\documentclass[aps,superscriptaddress,tightenlines,nofootinbib]{revtex4}
\usepackage{graphicx,amssymb,amsbsy,bm,amsmath}
\usepackage{hyperref}
\newcommand{\be}{\begin{equation}}
\newcommand{\ee}{\end{equation}}
\newcommand{\bea}{\begin{eqnarray}}
\newcommand{\eea}{\end{eqnarray}}

\newcommand{\OChi}{\ensuremath{\chi_{c}}}

\begin{document}
\title{New Inflation vs. Chaotic Inflation,  higher degree potentials
and the Reconstruction Program in light of WMAP3}
\author{D. Boyanovsky}
\email{boyan@pitt.edu} \affiliation{Department of Physics and
Astronomy, University of Pittsburgh, Pittsburgh, Pennsylvania 15260,
USA} \affiliation{Observatoire de Paris, LERMA. Laboratoire
Associ\'e au CNRS UMR 8112.
 \\61, Avenue de l'Observatoire, 75014 Paris, France.}
\affiliation{LPTHE, Universit\'e Pierre et Marie Curie (Paris VI) et
Denis Diderot (Paris VII), Laboratoire Associ\'e au CNRS UMR 7589,
Tour 24, 5\`eme. \'etage, 4, Place Jussieu, 75252 Paris, Cedex 05,
France}
\author{H. J. de Vega}
\email{devega@lpthe.jussieu.fr} \affiliation{LPTHE, Universit\'e
Pierre et Marie Curie (Paris VI) et Denis Diderot (Paris VII),
Laboratoire Associ\'e au CNRS UMR 7589, Tour 24, 5\`eme. \'etage, 4,
Place Jussieu, 75252 Paris, Cedex 05,
France}\affiliation{Observatoire de Paris, LERMA. Laboratoire
Associ\'e au CNRS UMR 8112.
 \\61, Avenue de l'Observatoire, 75014 Paris, France.}
\affiliation{Department of Physics and Astronomy, University of
Pittsburgh, Pittsburgh, Pennsylvania 15260, USA}
\author{C. M. Ho} \email{cmho@phyast.pitt.edu}
\affiliation{Department of Physics and Astronomy, University of
Pittsburgh, Pittsburgh, Pennsylvania 15260, USA}
\author{N. G. Sanchez}
\email{Norma.Sanchez@obspm.fr} \affiliation{Observatoire de Paris,
LERMA. Laboratoire Associ\'e au CNRS UMR 8112.
 \\61, Avenue de l'Observatoire, 75014 Paris, France.}
\date{\today}
\begin{abstract}
The CMB power spectra are studied for different  \emph{families} of
single field new and chaotic inflation models in the effective field
theory approach to inflation.  We implement a systematic expansion
in $ 1/N_e $ where $ N_e\sim 50 $ is the number of e-folds before
the end of inflation. We study the dependence of the observables ($
n_s, \; r $ and $ dn_s/d\ln k $) on the degree of the  potential ($
2 \, n $) and confront them to the WMAP3 and large scale structure
data: This shows in general that  fourth degree potentials ($ n=2 $)
provide the best fit to the data; the window of consistency with the
WMAP3 and LSS data narrows for growing $ n $. New inflation 
yields a good fit to the $ r $ and $ n_s $ data in a wide range of
field and parameter space. Small field inflation yields $ r<0.16 $
while large field inflation yields $ r>0.16 $ (for $ N_e=50 $). All
members of the new inflation family predict a small but negative
running $ -4 \, (n+1)\times 10^{-4} \leq  dn_s/d\ln k   \leq -2\times
10^{-4} $. (The values of $ r, \; n_s, \; dn_s/d\ln k $ for arbitrary
$ N_e $ follow by a simple rescaling from the $ N_e=50 $ values).
A reconstruction program is carried out suggesting quite
generally that for $ n_s $ consistent with the WMAP3 and LSS data
and $ r<0.1 $ the {\bf symmetry breaking scale} for new inflation is
$ |\phi_0| \sim 10~M_{Pl} $ while the {\bf field scale} at Hubble
crossing is $ |\phi_{c}| \sim M_{Pl} $.  The family of chaotic
models feature $ r \geq 0.16 $ (for $ N_e=50 $) and only a {\it
restricted subset} of chaotic models are consistent with the
combined WMAP3 bounds on $ r, \; n_s, \; dn_s/d\ln k $  with a
narrow window in field amplitude around $ |\phi_{c}| \sim 15~M_{Pl}
$. We conclude that a measurement of $ r<0.16 $ (for $ N_e =50 $)
distinctly rules out a large class of chaotic scenarios and favors
small field new inflationary models. As a general consequence, {\bf
new inflation} emerges more favoured than chaotic inflation.
\end{abstract}


\maketitle

\section{Introduction}
Inflation  provides a simple and robust mechanism  to solve several
outstanding problems of the standard Big Bang model
\cite{infla,libros} becoming a leading paradigm in cosmology.
Superhorizon quantum fluctuations amplified during inflation provide
an explanation of the origin of the temperature anisotropies in the
cosmic microwave background (CMB) and the seeds for large scale
structure formation\cite{fluc}, as well as of tensor perturbations
(primordial gravitational waves). Although there is a diversity of
inflationary models, most of them predict fairly generic features: a
gaussian, nearly scale invariant spectrum of (mostly) adiabatic
scalar and tensor primordial fluctuations \cite{fluc}. These
features provide an excellent fit to the highly precise data
provided by the Wilkinson Microwave Anisotropy Probe (WMAP)
\cite{WMAPa,peiris,WMAP3,WMAP3b} which begins to constrain
inflationary models.

The combination of CMB \cite{peiris,WMAP3} and large scale structure
data \cite{SDSS,2dF} yield fairly tight constraints for the two
dimensional marginalized contours of the tensor to scalar ratio $r$
and the scalar index $n_s$.  While $n_s=1$ was excluded at the $
95\%CL $ in \cite{2dF} a most notable result that stems from the
analysis of WMAP3 data is a confirmation that
  a scale invariant Harrison-Zeldovich
spectrum is excluded at the $3\sigma$ level \cite{WMAP3}. A
combination of data from WMAP3 and large scale surveys distinctly
favor $n_s <1$ \cite{sanchez}.  These latest bounds on the index of
the power spectrum of scalar perturbations, and emerging bounds on
the ratio of tensor to scalar fluctuations $ r $ begin to offer the
possibility to discriminate different inflationary models. For
example, the third year WMAP data disfavors the   predictions for
the scalar index and the    tensor to scalar ratio from a monomial
inflationary potential $ \lambda \; \phi^4 $ showing them to lie
outside the $ 3\sigma $ contour, but the simple monomial $ m^2 \;
\phi^2/2 $ yields a good fit to the data \cite{WMAP3} and predicts a
tensor to scalar ratio $ r\sim 0.16 $ within the range of
forthcoming CMB observations.

Current and future CMB observations in combination
with large scale structure surveys will yield tight constraints on
the inflationary models, this motivates the exploration of clear predictions
from the models and their confrontation with the data.

 In distinction with  the approach followed in
\cite{peiris,WMAP3,hoki}  or studies of specific
models\cite{salama}, or statistical analysis combined with WMAP3 and
LSS data\cite{kinkol,fipele}, we study the
predictions for the power spectra of scalar fluctuations and the
tensor to scalar ratio for \emph{families} of new and chaotic
inflationary models in the framework of the method presented in
ref.\cite{1sN}.   This method relies on the effective field theory
approach combined with a systematic expansion in $ 1/N_e $ where $ N_e
\sim 50 $ is the number of e-folds before the end of inflation. The
family of inflationary models that we study is characterized by
effective field theories with potentials of the form 
\bea 
V(\phi) &=&  V_0 -\frac12 \; m^2 \; \phi^2 + \frac{\lambda}{2 \, n}\;
\phi^{2 \, n} \quad , \quad {\rm broken ~ symmetry}\label{nuevito}\\
V(\phi) & = & \frac12 \; m^2 \; \phi^2 + \frac{\lambda}{2 \, n} \;
\phi^{2 \, n} \quad , \quad {\rm unbroken ~ symmetry} \; ,
\label{caoco}
\eea
with $ n=2,3,4\cdots $. For broken symmetry models
with potentials of the form (\ref{nuevito}) there are two distinct
regions: small and large field, corresponding to values of the
inflaton field smaller or larger than the symmetry breaking scale respectively.

We implement the systematic expansion in $ 1/N_e
$ where $ N_e\sim 50 $ is the number of e-folds before the end of
inflation when wavelengths of cosmological relevance crossed the
Hubble radius\cite{1sN}. The $ 1/N_e $ expansion is a powerful and
systematic tool that allows to re-cast the slow roll hierarchy as
expansion in powers of $ 1/N_e $ \cite{1sN}. This expansion
allows us to implement a   reconstruction program \cite{reco} which
yields the scale of the inflaton field when modes of cosmological
relevance today crossed the Hubble radius during inflation, and in
the case of new inflation models also yields the {\it scale of
symmetry breaking}.

\medskip

We study the dependence of the observables ($ n_s, \; r $ and $
dn_s/d\ln k $) on the degree of the inflaton potential ($ 2 \, n $)
for new and chaotic inflation and confront them with the WMAP3 data.
This study shows in general that  fourth degree potentials ($ n=2 $)
provide the best fit to the data. We find that new inflation fits
the data on an appreciable wider range of the parameters while
chaotic inflation does this in a much narrow range. Therefore,
amongst the families of inflationary models studied, new inflation
emerges as a leading contender in comparison with chaotic inflation.
The present analysis confirms  the statement that within the
framework of effective field theories with polynomial potentials,
new inflation is a preferred model reproducing the present data
\cite{ciri,pre,heclast}.

\medskip

{\bf Main results of this article:}

\begin{itemize}

\item{The region in inflaton field space which is consistent with the
marginalized WMAP3 data can be explored in an expansion in
$n_s-1+2/N_e$. }

 \item{ We find that the point $ n_s=1 -2/N_e= 0.96, \; r=8/N_e=0.16 $ which is in the region
allowed by the WMAP3 analysis\cite{WMAP3} belongs both to new inflation models
as a limiting point and to the simple chaotic inflation monomial, $ m^2 \, \phi^2/2 $.
(i) This point describes a
region in field and parameter space that separates small fields from
large fields,  and (ii) is a degeneracy point for the family of
models describing both chaotic and new inflation. }
\item{ For all members $ n = 2, \; 3, \; 4, ...$ of the new
inflation family, the small field region yields $ r<0.16 $ while the
large field region yields $ r>0.16 $. All members of the new inflation family
predict a small but negative running:
$$
-4 \; (n+1)\times 10^{-4} \leq  dn_s/d\ln k   \leq -2\times 10^{-4}  \; .
$$
This new inflation family features  a {\it large window} of
consistency with the WMAP and LSS data for $ n = 2 $ that narrows
for growing $ n $. If forthcoming data on tensor modes pinpoints the
tensor to scalar ratio to be $ r<0.1 $, we predict that the {\it
symmetry breaking scale} for these models is $ \phi_0 \sim
10\,M_{Pl} $ and that the scale of the field at which modes of
cosmological relevance today cross the Hubble radius is $ \phi_{c}
\sim M_{Pl} $.}

\item{Chaotic inflationary models all yield a tensor to scalar
ratio $ r \geq 0.16 $,  where the minimum value $ r=0.16 $
corresponds to small amplitude of the inflaton  and coincides with
the value obtained from the monomial $ m^2\phi^2/2 $.  The combined
marginalized data from WMAP3 \cite{WMAP3} yields a very small window
of field amplitude, around $|\phi_{c}|\sim 15~M_{Pl}$ within which
chaotic models are allowed by the data. These regions become
progressively smaller for larger $n$.  Some small regions in field
space consistent with the WMAP3 data feature peaks in the running of
the scalar index but in the region consistent with the WMAP3 data in
chaotic inflation the running is again negligible ($ \sim 10^{-3}
$).  If future observations determine a tensor to scalar ratio $
r<0.16 $, this by itself will  {\bf rule out} a large family of
chaotic inflationary models.}
\end{itemize}
\section{Effective field theory, slow roll and $ 1/N_e $ expansions}

In the absence of a fundamental microscopic description of
inflation, an effective field theory approach, when
combined with the slow roll expansion provides a robust paradigm for
inflation with predictive power. The reliability of the effective
field theory description hinges on a wide separation  between the
Hubble and Planck scales, and is validated by the bound from
temperature fluctuations $ H/M_{Pl} < 10^{-5} $ \cite{1sN}.

 The slow roll
expansion relies on the smallness of a hierarchy of the
dimensionless ratios \cite{libros,barrow,reco},
\be
\epsilon_v = \frac{M^2_{Pl}}{2} \; \left[\frac{V^{'}(\phi)}{V(\phi)} \right]^2
\quad , \quad \eta_v = M^2_{Pl} \; \frac{V^{''}(\phi)}{V(\phi)}
\quad , \quad
\label{etav} \xi_v = M^4_{Pl} \; \frac{V'(\phi) \;
V^{'''}(\phi)}{V^2(\phi)}\;. 
\ee
The effective field theory expansion in $H/M_{Pl}$ and the slow roll
expansion are independent, the latter can be interpreted as an
adiabatic expansion \cite{1sN} wherein the derivatives of the
inflationary potential are small.

The CMB data is consistently described within the slow roll
expansion with inflationary potentials of the form \cite{libros,1sN}
\be 
V(\phi) = M^4 \; v\left(\frac{\phi}{M_{Pl}}\right) \; .\label{potential} 
\ee 
Within the slow roll approximation the number
of e-folds before the end of inflation for which the value of the
field is $ \phi_{end} $ is given by 
\be 
N\left[\phi(t)\right] = -\frac{1}{M^2_{Pl}} \int_{\phi(t)}^{\phi_{end}} V(\phi)~
\frac{d\phi}{dV}~d\phi ~~ . 
\ee 
It proves convenient to introduce $ N_e  $ as the typical number of 
e-folds before the end  of inflation
during which cosmologically relevant wavelengths cross the Hubble
radius during inflation, and $ \phi_c $ as the value of the inflaton
field corresponding to $ N_e $ 
\be 
N_e = -\frac{1}{M^2_{Pl}} \int_{\phi_{c}}^{\phi_{end}} V(\phi)~ \frac{d\phi}{dV}~d\phi \,.
\label{Ne} 
\ee
The precise value of $ N_e $ is certainly near $ N_e=50 $ \cite{libros,Ne}.
We will take the value $ N_e=50 $ as a reference 
baseline value for numerical analysis, but from the explicit expressions
obtained in the systematic $ 1/N_e $ expansion below, it becomes a
simple rescaling to obtain results for arbitrary values of $ N_e $
[see eq. (\ref{etav2}) below].

\medskip

The form of the potential eq.(\ref{potential}) and the above
definition for the number of e-folds, suggests to introduce the
following rescaled field variable \cite{1sN} 
\be \label{chi} 
\phi = \sqrt{N_e} M_{Pl}\, \chi 
\ee 
where the rescaled field $\chi$ is
dimensionless. Furthermore, it is also convenient to scale $ N_e $
out of the potential and write 
\be 
V(\phi) = N_e \; M^4 \;  w(\chi) \; . 
\ee 
With this definition, eq. (\ref{Ne}) becomes 
\be \label{condi} 
1= - \int^{\chi_{end}}_{\chi_{c}} \frac{w(\chi)}{w'(\chi)}\,d\chi 
\ee 
where the prime stands for derivative with respect to $ \chi $,
$ \OChi $ is the value of $ \chi $ corresponding to $ N_e  $ e-folds before the end of
inflation, and $ \chi_{end} $ is the value of $ \chi $ at the end of inflation.

We emphasize that there is \emph{no} dependence on $ N_e $ in the
expression (\ref{condi}), therefore $ \chi_c,\chi_{end} $ only depend
on the coupling and the degree $ n $. The slow roll parameters
then become, 
\be \label{etav2} 
\epsilon_v = \frac{1}{2 \, N_e} \;
\left[\frac{w'(\chi_c)}{w(\chi_c)} \right]^2 \quad ,  \quad \eta_v =
\frac{1}{N_e} \; \frac{w''(\chi_c)}{w(\chi_c)} \;
  \quad ,  \quad
 \xi_v = \frac{1}{N^2_e} \; \frac{w'(\chi_c) \; w'''(\chi_c)}{w^2(\chi_c)}   \; .
\ee 
It is clear from eqs.(\ref{condi}) and  (\ref{etav2}) that
during the inflationary stage when wavelengths of cosmological
relevance cross the horizon, it follows that  $ w(\chi_c),
w'(\chi_c) \sim \mathcal{O}(1) $   leading to the   slow roll
expansion as a consistent  expansion in $ 1/N_e $ \cite{1sN}.

The connection between the slow roll expansion
and the expansion in $ 1/N_e $ becomes more explicit upon introducing
a \emph{stretched} dimensionless time variable $\tau$ and a
dimensionless Hubble parameter $h$ as follows \cite{1sN}
\be
\tau = \frac{M^2 \,t }{\sqrt{N_e} \; M_{Pl}} ~~;~~ h =
\frac{M_{Pl} \; H}{\sqrt{N_e} \; M^2} \label{slowvars} \; .
\ee
In terms of $ \tau , \; \chi $ and $ h $ the Friedmann equation and the
equation of motion for the inflaton become,
\bea
 h^2(\tau) = \frac{1}{3}\left[
\frac{1}{2 \, N_e}\left(\frac{d\chi}{d\tau}\right)^2+w(\chi)\right] \; ,
\label{fried} \cr \cr 
\frac{1}{N_e} \; \frac{d^2\chi}{d\tau^2}+ 3 \; h \;
\frac{d\chi}{d\tau} + w'(\chi) = 0 \; . \label{inflaeq}
\eea
This equation of motion can be solved in a systematic expansion in
$ 1/N_e $.  The definition (\ref{chi}) also makes manifest that $ \chi $
is a \emph{slowly varying field}, since a change  $ \Delta \phi \sim M_{Pl} $
in the inflaton field implies a \emph{small} change of the
dimensionless field $ \Delta \chi \sim 1/\sqrt{N_e} $.

In terms of these definitions, the CMB observables can be written
manifestly in terms of the $1/N_e$ expansion. The amplitude of
scalar perturbations is given by 
\be 
\Delta^2_{\mathcal{R}}  =
\frac{N^2_e}{12\pi^2} \; \left( \frac{M}{M_{Pl}}\right)^4\,
\frac{w^3(\chi_c)}{\left[ w'(\chi_c)\right]^2}  \; \, ,
\label{ampli} 
\ee 
and the spectral index $ n_s $, the ratio of
tensor to scalar perturbations $ r $ and  the running of $ n_s $
with the  wavevector $ dn_s / d\ln k $  become 
\bea 
n_s & = & 1-6 \;
\epsilon_v+ 2 \; \eta_v \label{ns} \quad , \quad r  =  16 \;
\epsilon_v \\ \cr \frac{dn_s}{d\ln k } & = & -\frac{2}{N^2_e}\left\{
\frac{w'(\chi_c)w'''(\chi_c)}{w^2(\chi_c)}+3 \left[
\frac{w'(\chi_c)}{w(\chi_c)}\right]^4-
4\,\frac{\big[w'(\chi_c)\big]^2\,w''(\chi_c)}{w^3(\chi_c)}\right\}
\label{dns} \; . 
\eea
The virtue of the $ 1/N_e $ expansion is that we can choose a
reference baseline value for $ N_e $, say $ N_e=50 $ for numerical study,
and use the scaling with $ N_e $ of the slow roll parameters given by
eqs.(\ref{etav2}), (\ref{ns}) and (\ref{dns}) to obtain their values
for arbitrary $ N_e $, namely 
\bea  
\epsilon_v[N_e]  & = &
\epsilon_v[50] \; \frac{50}{N_e} \quad , \quad
\eta_v[N_e] = \eta_v[50] \; \frac{50}{N_e} \quad , \quad
r[N_e] = r[50]  \; \frac{50}{N_e} \quad , \nonumber \\
 \frac{dn_s}{d\ln k }[N_e]  & = &    \frac{dn_s}{d\ln k }[50] \; \left(\frac{50}{N_e}\right)^2
\quad , \quad n_s[N_e] =  n_s[50]+\left(1- n_s[50]\right) \; \frac{N_e-50}{N_e}
   \; . \label{resca} 
\eea
In what follows we will take $N_e=50$ as   representative of the
cosmologically relevant case, however, the simple scaling relations
(\ref{resca})  allow  a straightforward extrapolation to other
values.

The combination of WMAP and SDSS (LRG) data yields \cite{WMAP3} 
\bea
&& n_s = 0.958\pm 0.016~~~~(\mathrm{assuming~} r=0 \mathrm{~with ~no~ running})\\
&& r < 0.28 ~ (95\% \, CL) ~~\mathrm{no ~running}
\\&& r < 0.67 ~ (95\% \, CL) ~~\mathrm{with ~ running}
\label{wmapvals} \; .
\eea
The running must be very small and of the order $ \mathcal{O}(1/N^2_e)\sim 10^{-3} $
in slow roll for generic potentials\cite{1sN}. Therefore, we can
safely consider $ dn_s/d\ln k =0 $ in our analysis. Figure 14 in
ref.\cite{WMAP3} and figure 2 in ref.\cite{heclast} show that the
preferred value of $n_s$ slowly grows with the preferred value of
$ r $ for $ r>0 $. We find approximately that 
\be 
\frac{\Delta n_s}{\Delta r} \simeq 0.12 \label{ley} 
\ee 
Therefore, for $ r\sim 0.1 $
the central value of $ n_s $ shifts from $ n_s = 0.958 ~(r=0) $ to $ n_s
= 0.97~ (r=0.1) $ as can be readily gleaned from the quoted figures
in these references.

As a simple example that
provides a guide post for comparison let us consider first the
monomial potential 
\be 
V(\phi) = \frac{\lambda}{2 \, n}  \, \phi^{2
\, n}\,. \label{quadpot } 
\ee 
The case $ n=1 $ yields a satisfactory
fit to the WMAP data \cite{peiris,WMAP3}.  For these potentials it
follows that, 
\be 
w(\chi) = \frac{\chi^{2 \, n}}{2 \, n}~~;~~ M^4 =
\lambda\,N_e^{n-1}\,M^{2 \, n}_{Pl}\, \,, \label{potchi} 
\ee
inflation ends at  $ \chi_{end} =0 $, and the value of the
dimensionless field $ \chi_c $ at $ \; N_e  $ e-folds before the end
of inflation is 
\be \label{chi50} 
|\chi_c| = 2 \; \sqrt{n} \; . 
\ee
These results lead to \cite{libros} 
\bea 
&&\epsilon_v   = \frac{n}{2
\, N_e} \quad , \quad \eta_v   = \frac{2 \, n-1}{2 \, N_e}\\&& n_s-1
= -\frac{n+1}{N_e} \label{nsmono} \quad , \quad
 r   =   \frac{8 \, n}{N_e} \quad , \quad \frac{dn_s}{d\ln k }   =
 -\frac{n+1}{N^2_e}\,.
\eea  
Taking  $ N_e=50 $ as a baseline, these yield 
\be 
n_s-1 = -2 \, (n+1)\times 10^{-2}~\Big(\frac{50}{N_e}\Big)  \quad , \quad r = 0.16
\, n ~\Big(\frac{50}{N_e}\Big) \quad , \quad
 \frac{dn_s}{d\ln k }=-4 \, (n+1)\times 10^{-4}~\Big(\frac{50}{N_e}\Big)^2 \; . \label{indicesmono}
\ee

\section{Family of models}

\begin{figure}[h]
\begin{center}
\includegraphics[height=3in,width=3in,keepaspectratio=true]{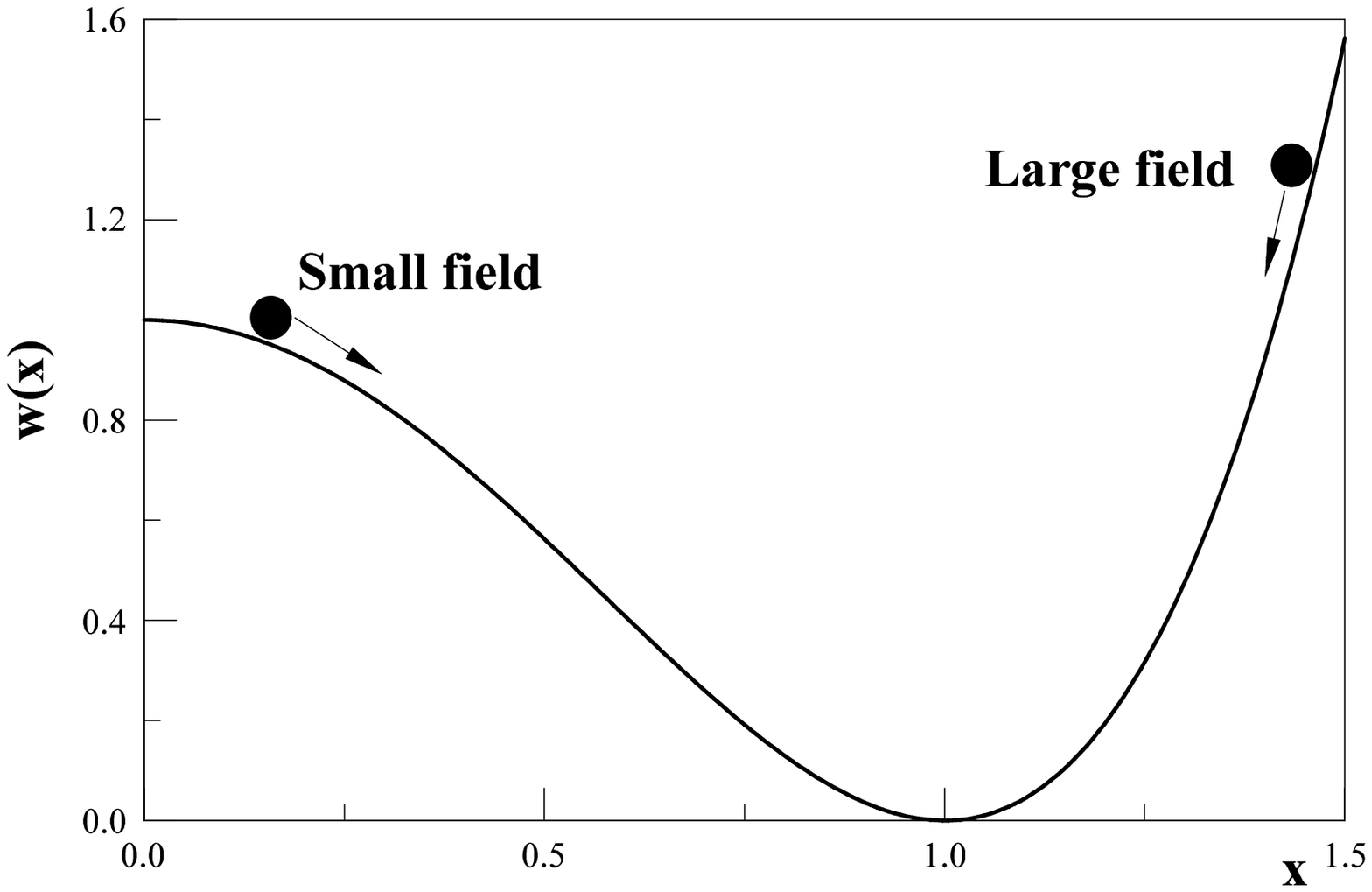}
\includegraphics[height=3in,width=3in,keepaspectratio=true]{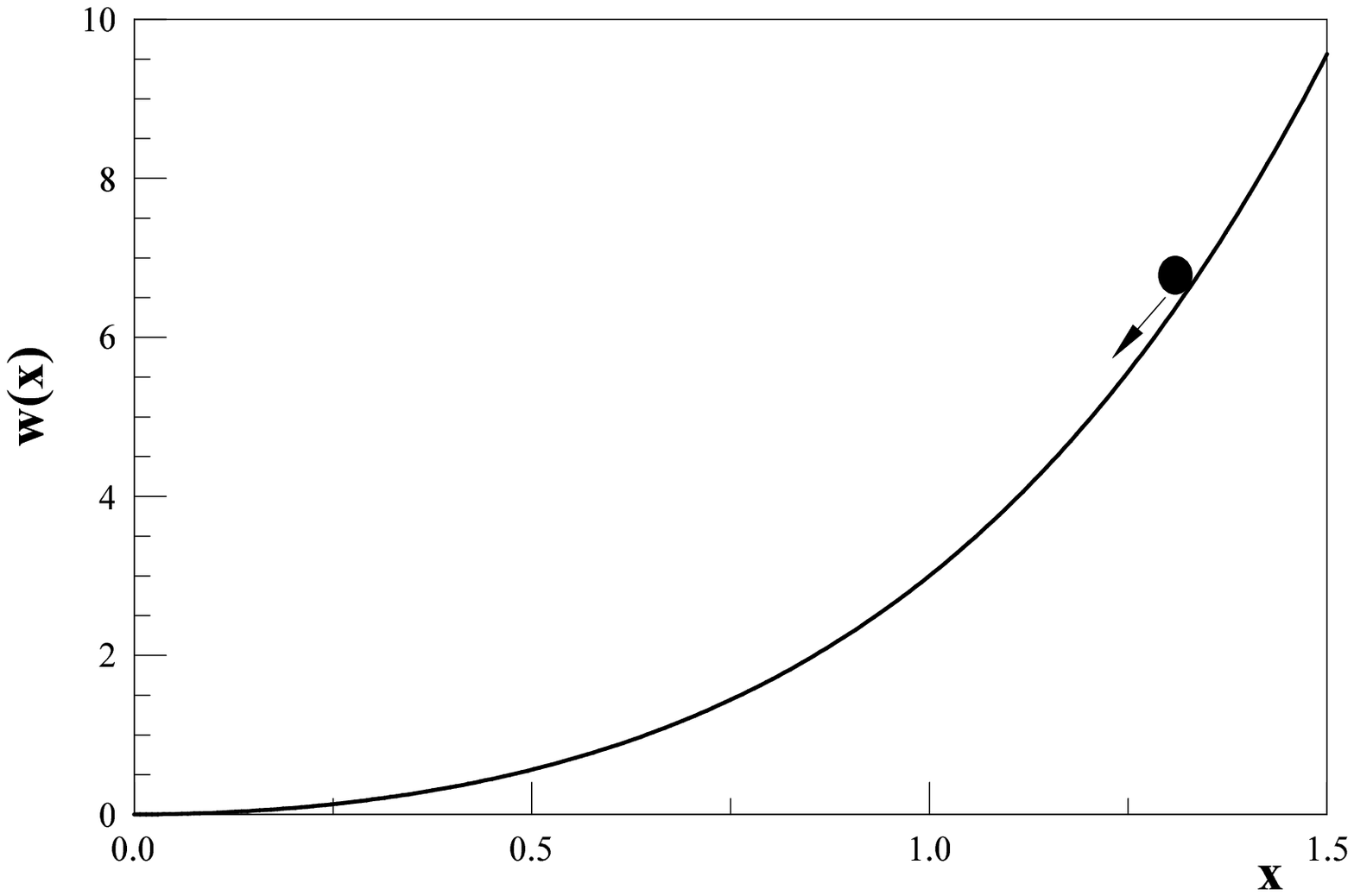}
\caption{Left panel: broken symmetry potential for  for $ n=2 $, small
and large field cases. Right panel: unbroken symmetry potential
for $ n=2 $, large field inflation. } \label{fig:newinflapot}
\end{center}
\end{figure}

We study the CMB observables $ n_s,r,dn_s/d\ln k $ for families of
new inflation and chaotic models determined by the following
inflationary potentials.
\bea
V(\phi) & = & V_0 -\frac{1}{2} \;
m^2 \; \phi^2 + \frac{\lambda}{2 \, n}\; \phi^{2 \, n} \quad , \quad {\rm
broken ~ symmetry} \label{rota}
\\ \cr V(\phi) & = & \frac{1}{2} \;  m^2 \; \phi^2 +
\frac{\lambda}{2 \, n} \; \phi^{2 \, n} \quad , \quad {\rm unbroken ~
symmetry} \; . \label{cao}
\eea
Upon introducing the rescaled field
$ \chi $ given by eq. (\ref{chi}) we find that these potentials can be
written as
\be
V(\phi) = N_e \;  m^2 \;  M^2_{Pl} \;  w(\chi)  \; ,
\ee
where we recognize that
\be
M^4 = m^2 \; M^2_{Pl} \; \, .
\ee
Then, the family of potentials eqs.(\ref{rota})-(\ref{cao}) are
\bea
&& w(\chi) = w_0-\frac{1}{2} \; \chi^2 + \frac{g}{2 \, n} \;
\chi^{2 \, n}\quad , \quad {\rm broken ~ symmetry} \label{neww}  \\ \cr
&& w(\chi) = \frac12 \; \chi^2 + \frac{g}{2 \, n} \; \chi^{2 \, n} \quad
, \quad {\rm unbroken ~ symmetry} \; \, .\label{newcao1}
 \eea
where
$ w_0 $ and $ g $ are dimensionless and related to $ V_0 $ and $
\lambda $ by
\be
V_0 = w_0 \;  N_e \;  m^2 \;  M^2_{Pl} \quad ,
\quad \lambda = \frac{m^2 \; g}{M^{2 \, n-2}_{Pl} \; N^{n-1}_e}   \; .
\ee
New inflation models described by the dimensionless potential
given by eq. (\ref{neww}) feature a  minimum at $ \chi_0 $ which is
the solution to  the following conditions
\be
w'(\chi_0)=w(\chi_0)=0
\; .\label{condchi0} \ee These conditions yield, \be \label{coupdef}
g = \frac{1}{\chi^{2 \, n-2}_0} \quad , \quad w_0 = \frac{\chi^2_0}{2 \, n}
\; \left(n-1\right)\; ,
\ee
$ \chi_0 $ determines the scale of
symmetry breaking $ \phi_0 $ of the inflaton potential upon the
rescaling eq.(\ref{chi}), namely
\be
\phi_0 = \sqrt{N_e} \; M_{Pl}
\;  \chi_0 \; \,. \label{fi0}
\ee
It is convenient to introduce the
dimensionless variable
\be
\label{defx} x = \frac{\chi}{\chi_0}
\ee
Then, from eqs.(\ref{coupdef}) and (\ref{defx}), the family of
inflation models eq.(\ref{neww})-(\ref{newcao1}) take the form
\bea
&&w(\chi) = \frac{\chi^2_0}{2 \, n}  \;   \left[ n \; (1-x^2) +
x^{2 \, n}-1\right] \quad , \quad {\rm broken ~ symmetry} \; ,
\label{newchi}\\
&&   w(\chi) = \frac{\chi^2_0}{2 \, n}  \left[n \;
x^2+x^{2 \, n}\right] \quad , \quad {\rm unbroken ~ symmetry} \; .
\label{caochi}
\eea
In terms of the variable $ x $, the small and
large field regions for the potential eq.(\ref{newchi}) correspond to $
x<1 $ and $ x>1 $, respectively. The form of the dimensionless
potentials $ w(x) $ for both families are depicted in fig.
\ref{fig:newinflapot}. The left panel in this figure shows the new
(small field) or chaotic (large field) behavior of the broken
symmetry potential eq.(\ref{newchi}) depending on whether the
initial value of the inflaton is smaller or larger than the minimum
of the potential $ x = 1 $.

\section{Broken Symmetry models.}

Inflation ends when the inflaton field arrives to the minimum of the
potential. For the new inflation family of models eq.(\ref{newchi})
inflation ends for  \be \chi_{end} = \chi_0 \; . \ee In terms of the
dimensionless variable $ x $, the condition eq.(\ref{condi}) becomes
\be 
\frac{2 \, n}{\chi^2_0} = I_n(X) \label{chi0I} \quad , \quad
X=\frac{\OChi}{\chi_0} \equiv x_{c} \; , \ee where \be \label{IXn}
I_n(X) = \int_{X}^1  \frac{dx}{x} \; \frac{n \; (1-x^2) + x^{2 \, n}
- 1 }{1-  x^{2 \, n-2}}= \int_{X}^1 \frac{n-\sum_{m=0}^{n-1}
x^{2m}}{\sum_{m=0}^{n-2} x^{2m}} \; \frac{dx}{x} \; . 
\ee 
This integral can be computed in closed form in terms of hypergeometric
functions \cite{gr} which can be reduced to a finite sum of
elementary functions\cite{pru}.

For a fixed given value of $ X $, the value of $ \chi_0 $ and
therefore of the dimensionless coupling $ g $ is determined by the
equation (\ref{chi0I}). Once we obtain this value, the CMB
observables eqs. (\ref{ns})-(\ref{dns}) are obtained by evaluating
the derivatives of $ w(\chi) $ at the value $ \chi = \OChi = \chi_0
\; X $ with the corresponding value of the coupling $ g $. Thus, a
study of the range of possible values for $ n_s, \; r, \; dn_s/d\ln
k $ is carried out by exploring the relationship between these
spectral indices as a function of $ X $. 
For this study we choose the baseline value $ N_e=50 $ from which the
indices can be obtained for arbitrary value of $N_e$ by the relation
(\ref{resca}).

While the dependence of $ \chi_0 $ and $ g $ upon the variable $ X $ must
in general be studied numerically, their behavior in the
relevant limits, $ X \rightarrow (0, \; 1) $ for small field inflation and $ X >> 1 $
for large field inflation can be derived  from eqs.(\ref{chi0I})-(\ref{IXn}).

For small field inflation and  $ X \rightarrow 0 $    the lower
limit of the integral dominates leading to
\be
\chi^2_0 \buildrel{X
\to 0}\over= \frac{2 \, n}{n-1} \; \frac1{\log\frac1{X}} \quad ,
\quad g \buildrel{X \to 0}\over= \left( \frac{n-1}{2 \, n} \;
\log\frac1{X}\right)^{n-1} \; , \label{smallX}
\ee
thus, as $ X \rightarrow 0 $ these are \emph{strongly coupled theories}. This
result has a clear and simple interpretation: for $ N_e =50 $ to be
the number of e-folds between $ x=X $ and $ x=1 $ the coupling $ g $
must be large and the potential must be steep, otherwise there would
be many more e-folds in such interval.

For small field inflation and $ X \rightarrow 1^{-} $ the integral $ I_n(X) $
obviously vanishes and
\be\label{Xten1}
\chi^2_0 \buildrel{X \to  1^{-}}\over=\left( \frac2{1-X}\right)^2
\left[ 1 + \frac{2\, n - 1}9 \; (X-1) + {\cal O}(X-1)^2 \right]
\quad , \quad g \buildrel{X \to  1^{-}}\over= =\left[\frac12 \; (1-X)\right]^{2\,(n-1)}
\rightarrow 0\; ,
\ee
thus, as $ X \rightarrow 1^{-} $, these are a \emph{weakly coupled family of models}.

\medskip

\begin{figure}[ht!]
\begin{center}
\includegraphics[height=3.5in,width=3.5in,keepaspectratio=true]{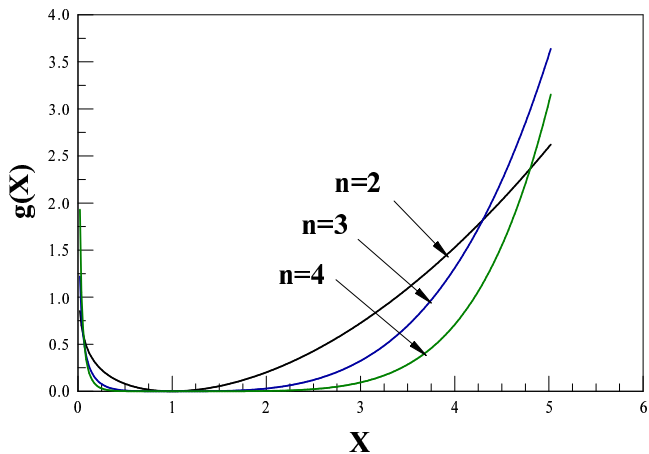}
\caption{The coupling $ g $ as a function of $ X $, for the degrees
of the new inflation potential $ n=2, \; 3, \; 4 $. For $ X \to 1 ,
\; g $ vanishes as $ \left[\frac{1-X}2\right]^{2\,(n-1)} $. The
point $ X = 1, \; g = 0 $ corresponds to the monomial $ m^2 \;
\phi^2/2 $. $ g $ increases both for $ X \to 0 $ and for large $ X $
as, $ \left( \log\frac1{X}\right)^{n-1} $ and $ \left( \frac{X^2}{4
\, n} \right)^{n-1} $, respectively.} \label{fig:couplingsni}
\end{center}
\end{figure}

For large field inflation and $ X \gg 1 $, the integral $ I_n(X) $ is dominated by the
term with the highest power, namely $ x^{2 \, n} $, leading to the behavior
\be\label{Xgran}
\chi^2_0 \buildrel{X \gg 1}\over= \frac{4 \, n}{X^2}
\quad , \quad g \buildrel{X \gg 1}\over= \left( \frac{X^2}{4 \, n} \right)^{n-1} \; ,
\ee
which leads to a strongly coupled regime.
The results of a numerical analysis are depicted in fig. \ref{fig:couplingsni}.

Before we proceed with a numerical study of the CMB indices and the
tensor to scalar ratio, we can extract interesting and relevant
information by focusing on the region $ X \sim 1 $ which as
discussed above corresponds to a weakly coupled family for broken
symmetry potentials. This is the region near the \emph{minimum} of
the potential and the integral $ I_n(X) $ can be evaluated simply by
expanding $ w(\chi) $ and its derivative near the minimum. To
leading order in $ (X-1) $ the condition (\ref{chi0I}) leads to
eq.(\ref{Xten1}) and 
\be \label{quadcond} 
(\chi_c-\chi_0)^2 = 4
\quad {\rm or} \quad |\chi_c-\chi_0| = 2 \; . 
\ee 
This is precisely
eq.(\ref{chi50}) for $ n = 1 $ upon the shift $ \chi_c \to
\chi_c-\chi_0 $. Namely, eq.(\ref{quadcond}) is the condition
eq.(\ref{chi50}) for the quadratic monomial potential with minimum
at $ \chi = \chi_0 $ instead of $ \chi = 0 $ as in
eq.(\ref{potchi}). This is clearly a consequence of the fact that
near the minimum $ X = 1 $ the potential is quadratic, therefore for
$ X \sim 1 $ the quadratic monomial is an excellent approximation to
the family of higher degree potentials and more so because $ g \sim
0 $. For $ X\sim 1 $   we find to leading order in $ (X-1) $ the
values: 
\be 
\epsilon_v =\eta_v = 0.01 ~\Big(\frac{50}{N_e}\Big)
\quad , \quad n_s = 0.96+0.04~\Big(\frac{N_e-50}{N_e}\Big) \quad ,
\quad r = 0.16~\Big(\frac{50}{N_e}\Big) \quad , \quad
\frac{dn_s}{d\ln k } = - 0.0008~\Big(\frac{50}{N_e}\Big)^2 \; .
\label{quadvals} 
\ee 
The fact that the potential eq.(\ref{newchi})
is quadratic around the minimum $ X = 1 $ explains why we have in
this limit identical results for new inflation with the potential
eq.(\ref{newchi}) and chaotic inflation with the monomial
potential $ m^2 \; \phi^2/2 $.

As observed in \cite{WMAP3} these values of $ r, \; n_s $ for $N_e
\sim 50$ yield a good fit to the available CMB data.

\begin{figure}[h]
\begin{center}
\includegraphics[height=3 in,width=3 in,keepaspectratio=true]{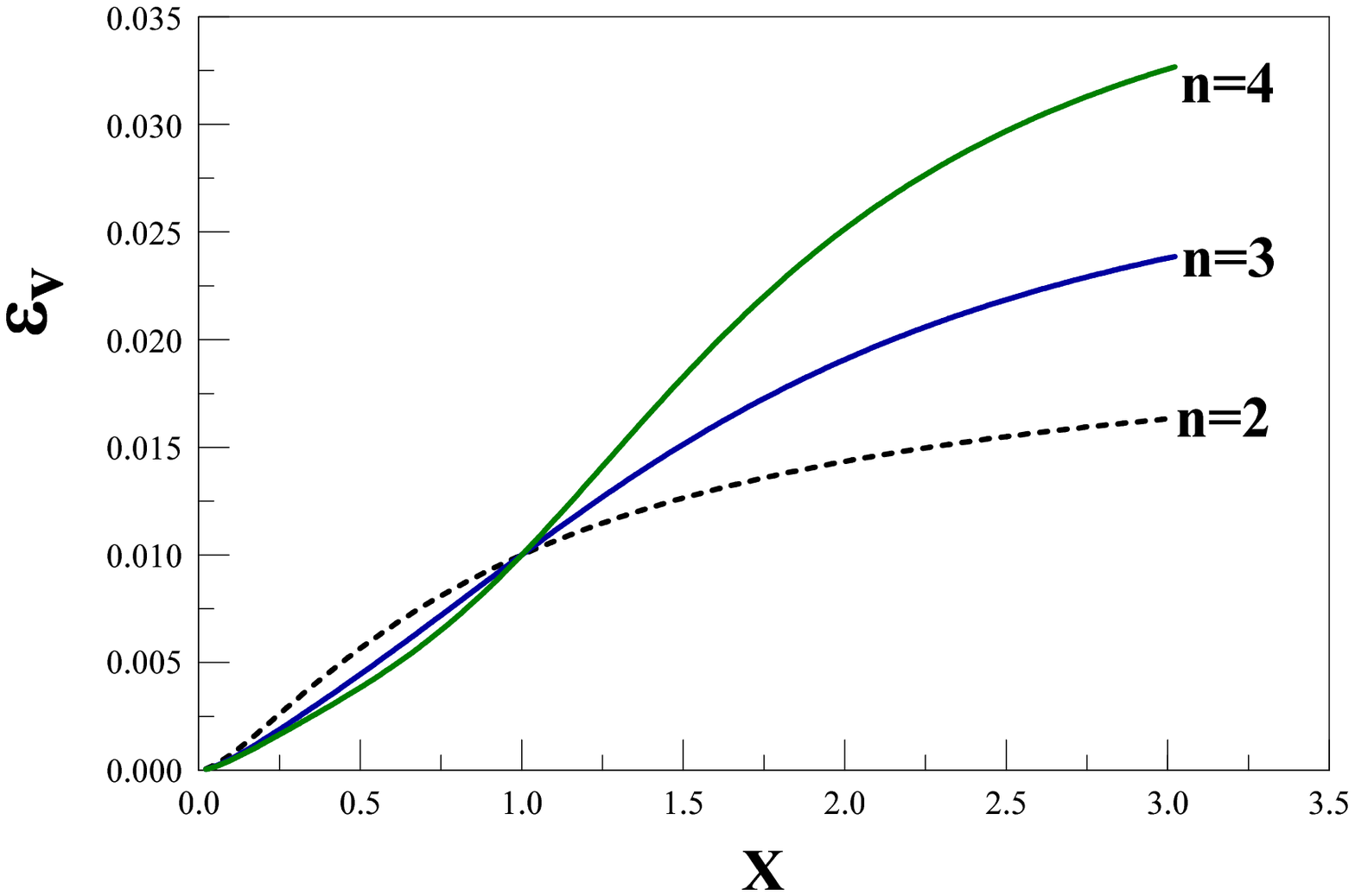}
\includegraphics[height=3in,width=3in,keepaspectratio=true]{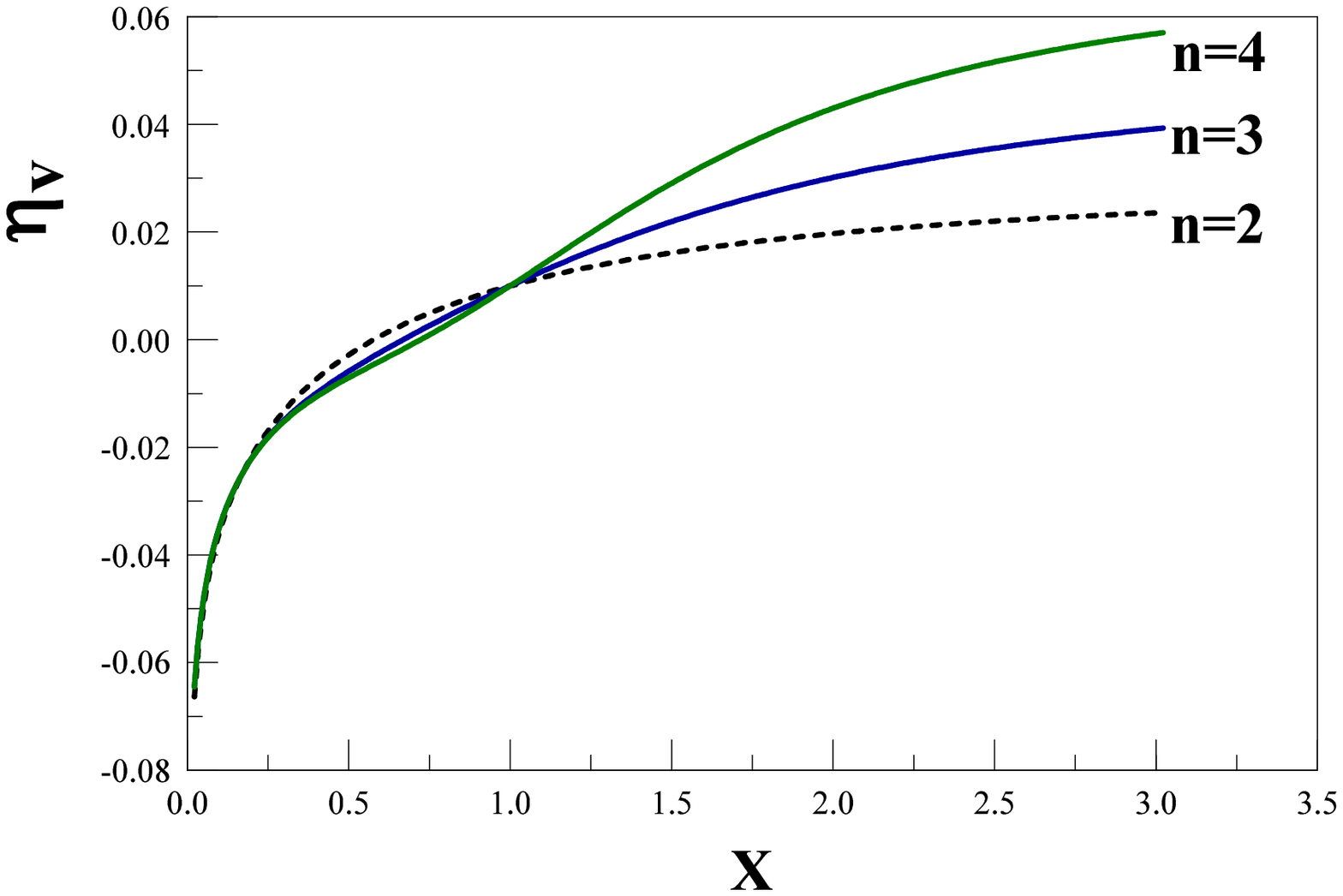}
\caption{Slow roll parameters as a function of $ X $ for $ N_e=50 $.
Left panel $ \epsilon_v $, right panel $ \eta_v $, for new inflation
with the degrees of the potential $ n=2,3,4 $. The results for
arbitrary values of $ N_e $ are obtained by multiplying by the factor
$ \frac{50}{N_e} $.} \label{fig:epsilonv}
\end{center}
\end{figure}

\begin{figure}[h!]
\begin{center}
\includegraphics[height=3.5in,width=3.5in,keepaspectratio=true]{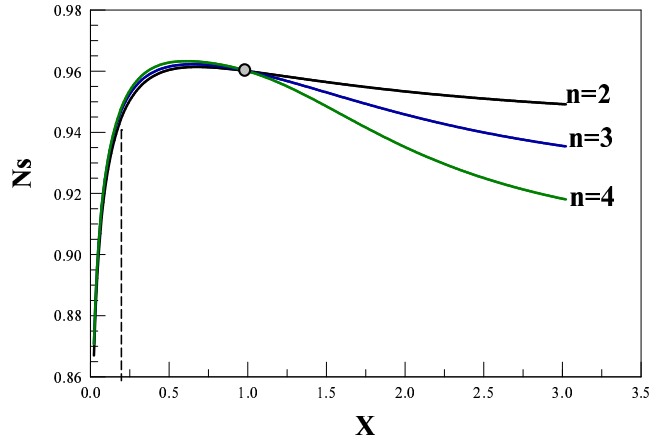}
\caption{Scalar spectral index $ n_s $ for the degrees of the
potential $ n=2,3,4 $ for new inflation with $ N_e=50 $.  The
vertical line delimits the smallest value of $n_s$ (for
$ r=0 $) \cite{WMAP3}. The grey dot at $ n_s=0.96, \; X=1 $
corresponds to the value for the monomial potential $ n = 1 , \; m^2
\; \phi^2/2 $. Notice that the small field behavior is $ n $
independent. For arbitrary $ N_e $ the result follows directly 
from the $ N_e=50 $ value by using eq.(\ref{resca}).} \label{fig:NsNI}
\end{center}
\end{figure}

\begin{figure}[h!]
 \begin{center}
 \includegraphics[height=3.5 in,width=3.5 in,keepaspectratio=true]{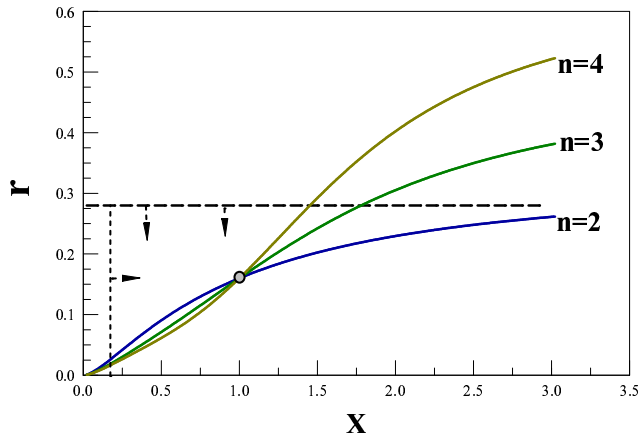}
 \caption{Tensor to scalar ratio $ r $ vs. $ X $
  for the degrees of the potential $ n=2, \; 3, \; 4 $ for
new inflation with $ N_e=50 $.
The horizontal dashed line corresponds to the upper limit $ r=0.28~(95\% CL) $
  from WMAP3 without running. The vertical dashed line
determines the minimum value of $ X , \; X \sim 0.2 $, consistent
with the WMAP limits for $ n_s $ as in fig. \ref{fig:NsNI}.
The grey dot at $ X=1, \; r=0.16 $ corresponds to the value for the
monomial potential $ m^2 \; \phi^2/2 $. The small field limit is
nearly independent of $ n $. For arbitrary $ N_e $ the result follows directly 
from the $ N_e=50 $ value by using eq. (\ref{resca}).}
 \label{fig:rni}
 \end{center}
 \end{figure}

\begin{figure}[h!]
 \begin{center}
 \includegraphics[height=3.5 in,width=3.5 in,keepaspectratio=true]{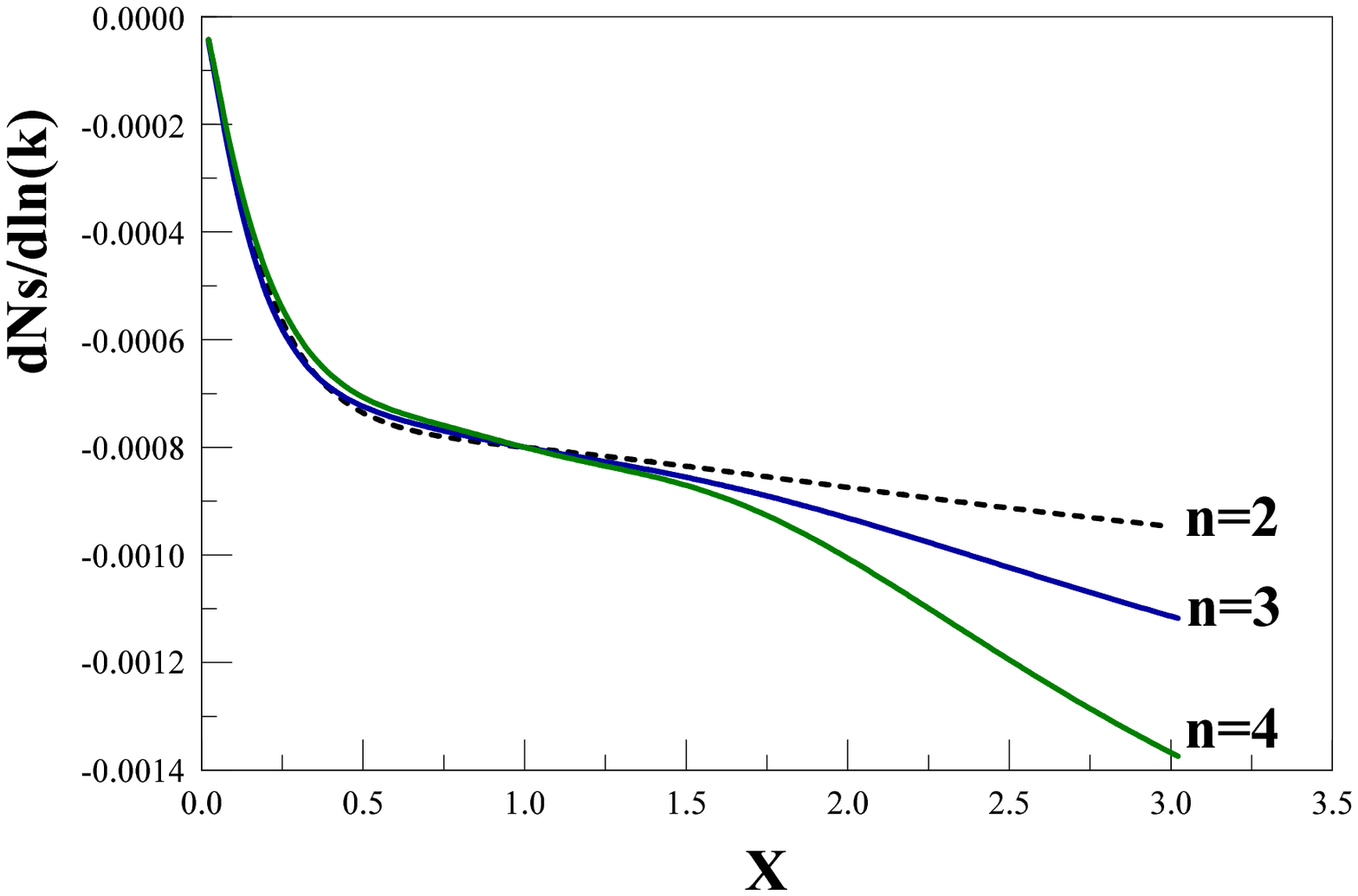}
 \caption{Running of the scalar index $ dn_s/d\ln k $ vs. $ X $ for
the degrees of the potential $ n=2,3,4 $ respectively for new
inflation with $ N_e=50 $. The small field behavior is independent
of $ n $. For arbitrary $ N_e $ the result  follows directly 
from the $ N_e=50 $ value by using eq. (\ref{resca}).}
 \label{fig:dnsni}
 \end{center}
 \end{figure}

The results of the numerical analysis for $ \epsilon_v, \; \eta_v,
\; n_s, \; r $ and $ dn_s/d\ln k $   for the baseline value $N_e=50$
are depicted in fig. \ref{fig:epsilonv}, \ref{fig:NsNI},
\ref{fig:rni} and \ref{fig:dnsni} respectively.

 The vertical dashed line  in fig. \ref{fig:NsNI} at $ X \sim 0.2 $
determines the lower limit of $ X $ for which $ n_s $ is consistent
with the WMAP data for $r=0$. For large values of $ X , \; n_s $
approaches asymptotically the values for the monomial potentials $
\phi^{2 \, n} $ given by eq. (\ref{indicesmono}). For the larger
degrees $ n $, the asymptotic behavior of  $ n_s $ and $ r $ settles
at larger values of $ X $, this is a consequence of the larger
region in which the coupling is small as observed in fig.
\ref{fig:couplingsni} for larger  degrees $ n $. The horizontal
dashed lines with vertical downward arrows in fig. \ref{fig:rni}
determines the upper bound from WMAP \cite{WMAP3} given by eq.
(\ref{wmapvals}) \emph{without} running, since from fig.
\ref{fig:dnsni} the values of $ dn_s/d\ln k $ for these models are
negligible. The vertical dashed line with the right-pointing arrow
in fig. \ref{fig:rni} determines the values of $ X $ for which $ n_s
$ are consistent with the WMAP data (see also fig. \ref{fig:NsNI}).

\medskip

From these figures we see that unlike the case of a pure monomial potential
$ \lambda \, \phi^{2 \, n} $ with $ n\geq 2 $, there is a {\bf large} region of field
space within which the new inflation models given by eq.(\ref{rota}) are {\bf consistent}
with the bounds from marginalized WMAP3 data and the combined WMAP3 + LSS
data \cite{WMAP3}.

\medskip

\begin{figure}[h]
 \begin{center}
 \includegraphics[height=3.5 in,width=3.5 in,keepaspectratio=true]{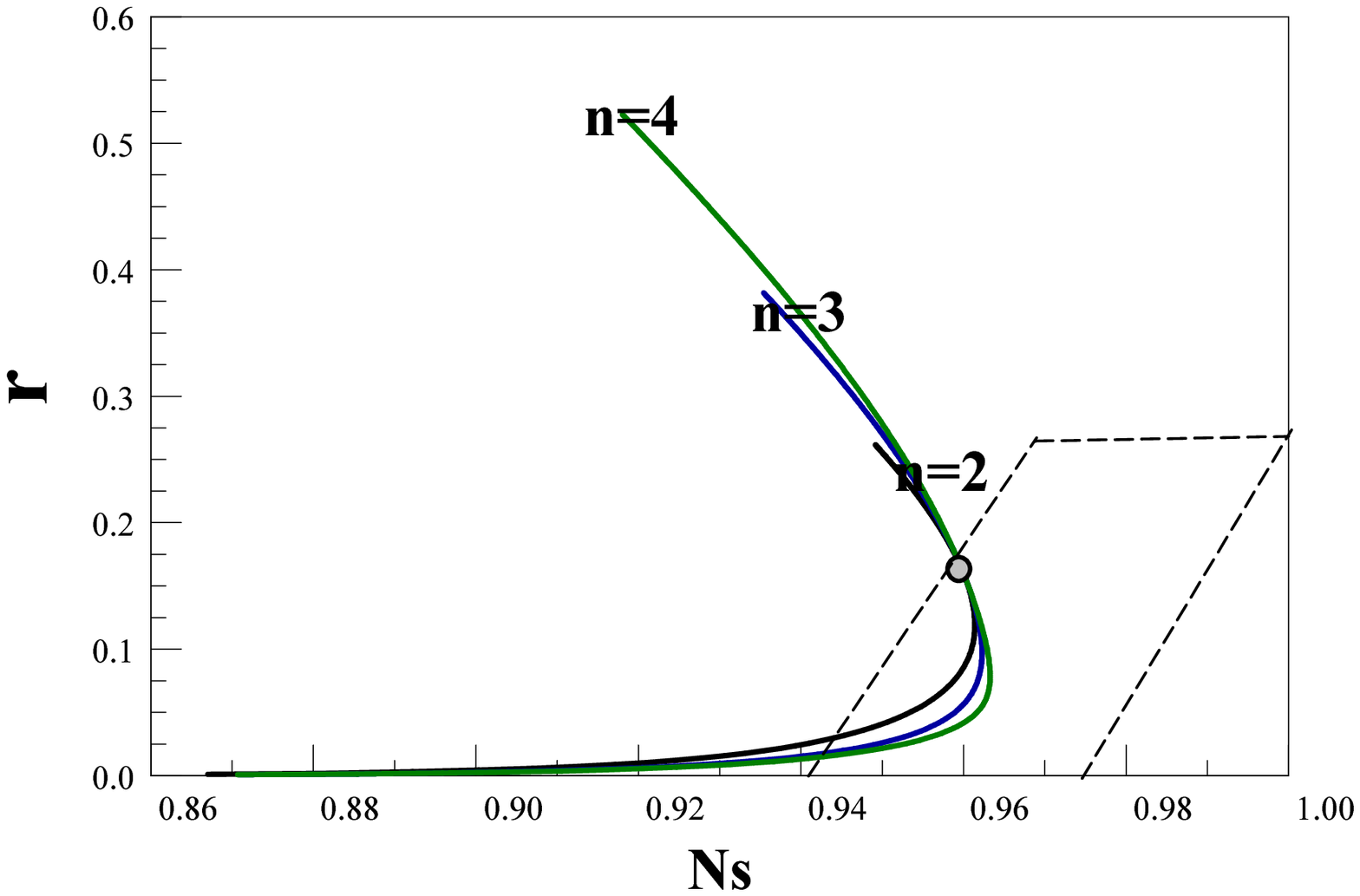}
 \caption{Tensor to scalar ratio $ r $ vs. $ n_s $
  for degrees of the potential $ n=2, \, 3, \, 4 $ respectively for new inflation with $ N_e=50 $.
 $ r $ turns to be a {\bf double-valued} function
of  $ n_s $ exhibiting a maximum value for $ n_s $.
The values inside the box between the dashed lines
  correspond to the WMAP3 marginalized region of the
$(n_s,r)$ plane with $ (95\% CL): \; r < 0.28, \;  0.942+0.12 ~ r
\leq n_s \leq 0.974+0.12 ~ r $, see eq.(\ref{ley}). The grey dot
corresponds to the values for the monomial potential $ m^2\phi^2/2 $
and the value $ X=1: \; r = 0.16, \; n_s = 0.96 $.}
 \label{fig:rvnsni}
 \end{center}
 \end{figure}

Fig. \ref{fig:rvnsni} displays $ r $ vs $ n_s $ for the values $
n=2,3,4 $ in new inflation and indicate the trend with $n$. 
While $ r $ is a monotonically increasing function of $ X, \; n_s $ features a
\emph{maximum} as a function of $ X $, hence $ r $ becomes a {\bf
double-valued} function of $ n_s $. The grey
dot at $ r=0.16, \; n_s =0.96 $ corresponds to the monomial
potential $ m^2 \; \phi^2/2 $ for $ N_e=50 $. Values below the grey
dot along the curve in fig. \ref{fig:rvnsni} correspond to small
fields $ X<1 $ while values above it correspond to large fields $
X>1 $. We see from figs. \ref{fig:rni} and \ref{fig:rvnsni} that
\emph{large fields systematically lead to larger values of}  $ r $.
Models that  fit the WMAP data to $ 95\%~CL $ are within the tilted
box in fig. \ref{fig:rvnsni}.  The tilt accounts for the growth of the preferred value of
$ n_s $ with $ r $ \cite{WMAP3} according to eq. (\ref{ley}). 

\begin{figure}[h]
 \begin{center}
 \includegraphics[height=3.5 in,width=3.5 in,keepaspectratio=true]{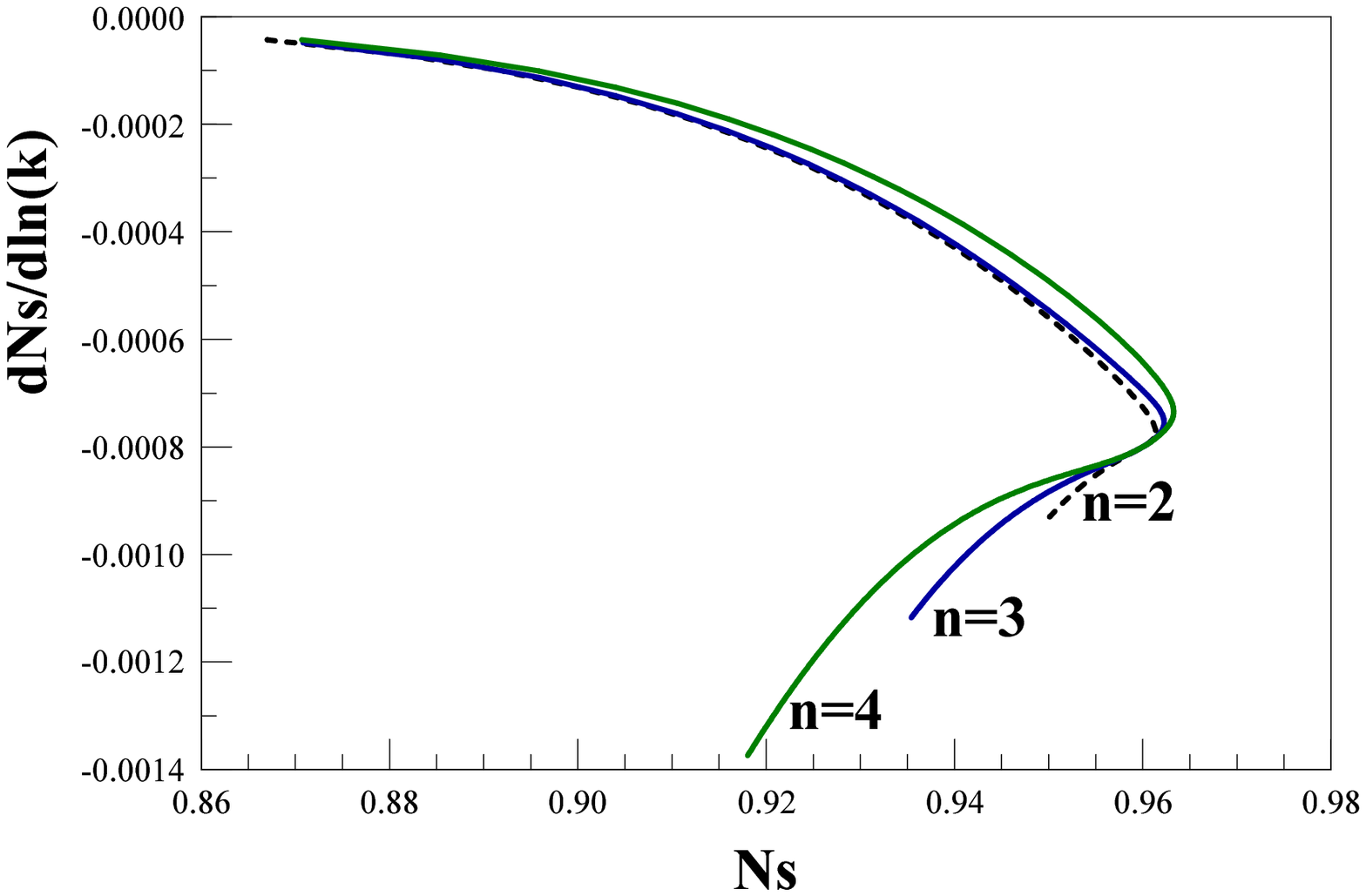}
 \caption{Running of the scalar index $ dn_s/d\ln k $ vs. $ n_s $ for degrees of the potential
   $ n=2, \, 3, \, 4 $ respectively for new inflation with $ N_e=50 $.
The values for arbitrary $ N_e $ follow directly 
from the $ N_e=50 $ value by using eq. (\ref{resca}).}
 \label{fig:ndnsni}
 \end{center}
 \end{figure}

\medskip

 Fig. \ref{fig:ndnsni} displays the running of the scalar index
 vs. $ n_s $ for the different members of the family of new inflation
showing clearly that running is all but negligible in the entire range of values
 consistent with the WMAP data. This was expected since the running in
slow-roll is of the order $ \sim \frac1{N_e^2} \simeq 4 \times 10^{-4} $
[see eq.(\ref{dns})] \cite{1sN}.

We note that $ dn_s/d\ln k $ is a
 monotonically \emph{decreasing} function of $ X $ approaching
 asymptotically the values for the monomials $ \phi^{2 \, n} $ given by
 eq. (\ref{indicesmono}). $ X \sim  0.2$ which is the {\it minimum
 value} of $ X $ consistent with the bounds from WMAP on $ n_s $ (see
 fig. \ref{fig:NsNI}).

 \subsection{Field reconstruction }

 The above analysis suggests to study the \emph{inverse problem},
 namely, for a given member of the family labeled by $ n $, we may ask
 what is the value  $ \phi_{c} $ of the field at Hubble crossing and what is the scale
 $ \phi_{0} $ of symmetry breaking of the potential which are consistent with the
 CMB+LSS data. This is tantamount to the program of reconstruction
 of the inflaton potential advocated in ref. \cite{reco}  and is
 achieved as follows: eq.(\ref{chi0I}) yields
 $ \chi_0=\chi_0[X] $ from which we obtain $ \chi_c = \chi_0 \; X $.
These results are then input into the
 expression for $n_s$ by evaluating the potential $ w(\chi) $ and its
derivatives at the value of
 $ \chi_c $. This yields $ n_s=n_s[\chi_c] $ which is then
 inverted to obtain $ \chi_c= \chi_c[n_s] $ and thus $ \phi_{c} $.

In the region $ X \sim 1 $ corresponding to the weakly coupled case, this
 reconstruction program can be carried out as a systematic series in
\be \label{delta}
\Delta \equiv  X -1 = \frac{\OChi}{\chi_0}-1   \; ,
\ee
by expanding the inflationary potential and its derivatives
 in a power series in $ x $ around $ x=1 $ in the integrand of $ I_n(X) $ [eq.
 (\ref{IXn})]. For $ X=1 $ the value of the scalar index $ n_s $ is
 determined by the simple monomial $ m^2 \; \phi^2/2 $ which from eq.
 (\ref{nsmono}) for $ n=1 $ is given by $ n_s-1 = -2/N_e $.
Therefore, in terms of $ n_s $, the actual expansion parameter is $ n_s-1+2/N_e $.

We obtain $ n_s $ to first order in $ \Delta $ from
eqs.(\ref{etav2}), (\ref{ns}), (\ref{newchi}), (\ref{Xten1})  and
(\ref{delta}) with the result,
$$
 n_s-1 = -\frac2{N_e}\left[ 1 + \frac{2 \, n - 1}{18} \; \Delta + {\cal O}(\Delta^2)\right]
$$
then, by inverting this equation we find:
\be
\Delta(n_s,n) = X-1 = -\frac{9
 N_e}{2 \, n-1}\left( n_s-1+\frac{2}{N_e}\right) +
\mathcal{O}\left(\left[ n_s-1+\frac{2}{N_e}\right]^2\right) \; , \label{delns}
 \ee
and from eqs. (\ref{Xten1}) and (\ref{delns}) we find,
\be
\chi_0(n_s,n)  =
 \frac{2 \; (2 \, n-1)}{9 \, N_e \left| n_s-1+\frac{2}{N_e}\right|}\left[ 1- \frac{N_e}{2}
\left( n_s-1+\frac{2}{N_e}\right)  \right] +
\mathcal{O}\left(\left[ n_s-1+\frac{2}{N_e}\right]\right)
\label{chi0s}
\ee
The leading order ($ \propto 1/\Delta $) of this result for $ \chi_0(n_s,n) $
can be simply cast as eq.(\ref{quadcond}):
this is recognized as the condition to have $ 50 $ e-folds for the quadratic
 monomial centered in the broken symmetry minimum [see discussion below eq.
 (\ref{quadcond})].

\medskip

Finally, the value of the (dimensionless) field $ \chi_c $ at Hubble
crossing is determined
 from $ \chi_c(n_s,n) = \chi_0 \; [1+\Delta(n_s,n)] $ from which we obtain
 \be
\chi_c = \frac{2 \; (2 \, n-1)}{9 \, N_e \left|
n_s-1+\frac{2}{N_e}\right|}\left[ 1- \frac{(2 \, n+17) \; N_e}{2 \;
(2 \, n-1)}\left( n_s-1+\frac{2}{N_e}\right)
 \right] + \mathcal{O}\left(\left[ n_s-1+\frac{2}{N_e}\right]\right) \; .
 \label{chi50s}
\ee
The coupling constant $ g $ can be also expressed in terms of $ n_s $
in this regime with the result,
$$
g = \left[ \frac{9 \, N_e \left| n_s-1+\frac{2}{N_e}\right|}{2 \; (2 \, n-1)}\right]^{2 \, n - 2}\to 0 \; ,
$$
which exhibits the weak coupling character of this limit.

\begin{figure}[h]
 \begin{center}
 \includegraphics[height=2.5in,width=2.5in,keepaspectratio=true]{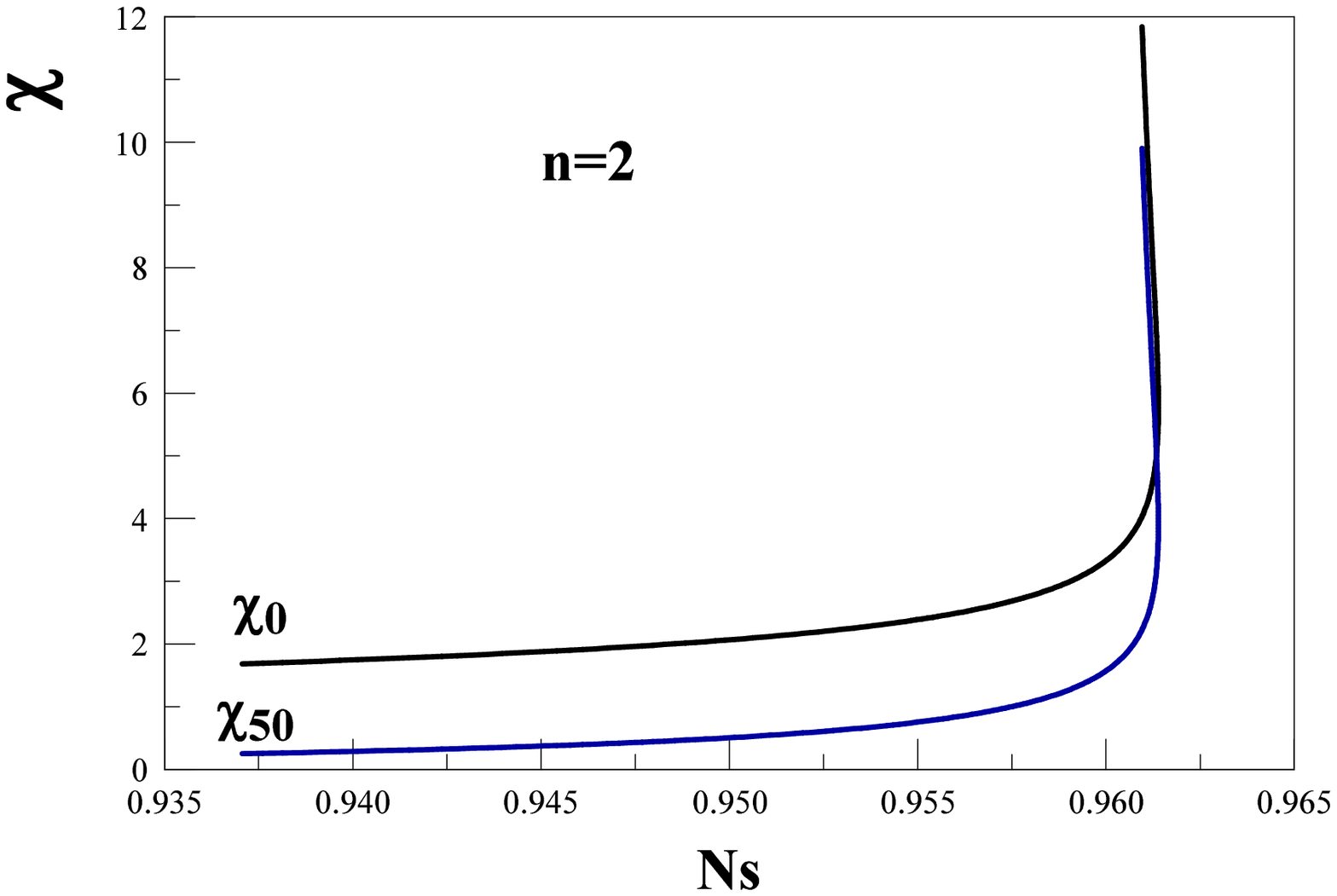}
 \includegraphics[height=2.5in,width=2.5in,keepaspectratio=true]{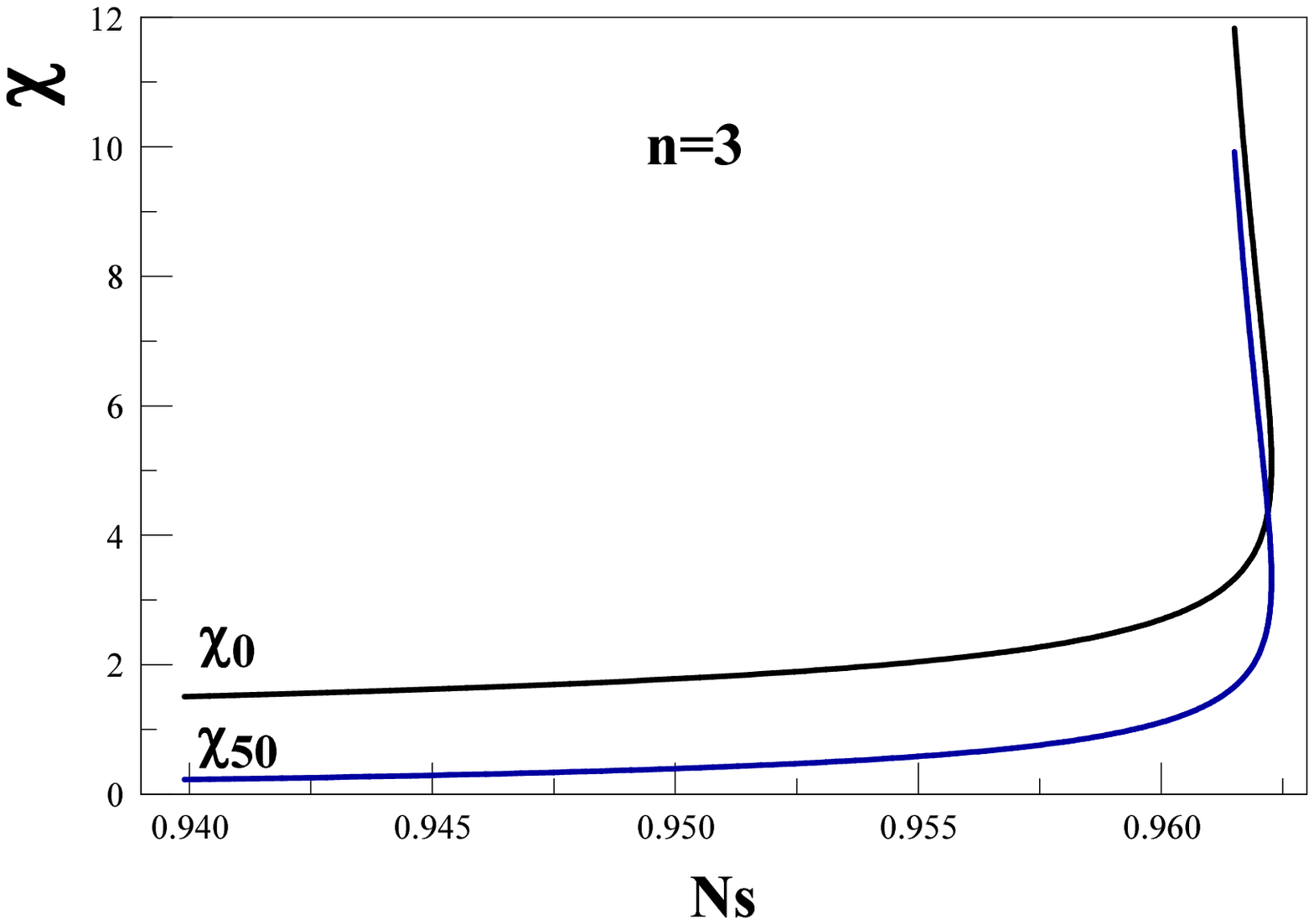}
 \includegraphics[height=2.5in,width=2.5in,keepaspectratio=true]{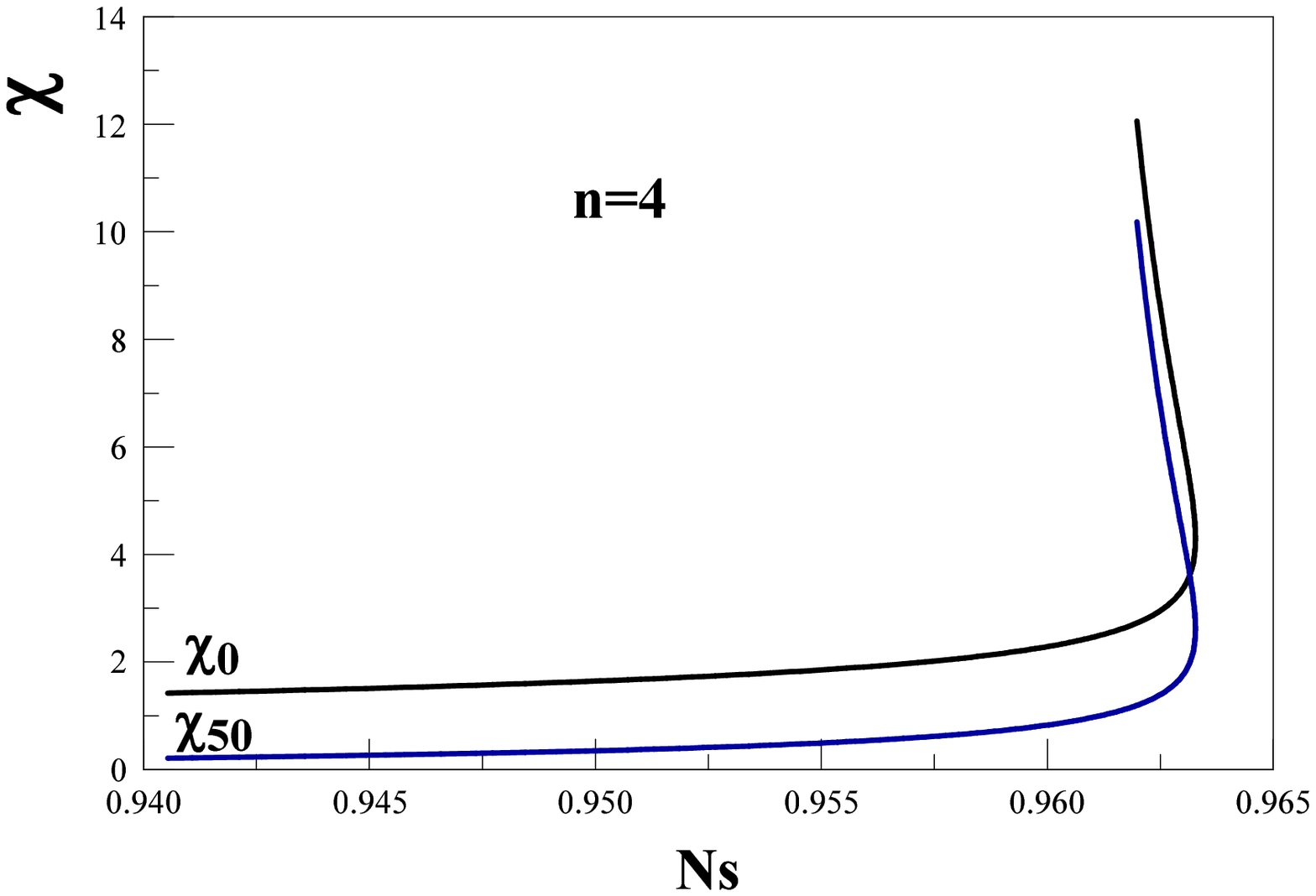}
 \caption{Reconstruction program for broken symmetry potentials with $ N_e=50 $,
small field case $ X<1 . \; \chi_{50}\equiv \chi_c $ and $ \chi_0$
vs. $ n_s $ for degrees of the potential
   $ n=2,3,4 $, respectively. These values of $ \chi_c, \; n_s $ correspond to the region $ r<0.16 $.}
 \label{fig:chinsniSF}
 \end{center}
 \end{figure}

\begin{figure}[h]
 \begin{center}
 \includegraphics[height=2.5in,width=2.5in,keepaspectratio=true]{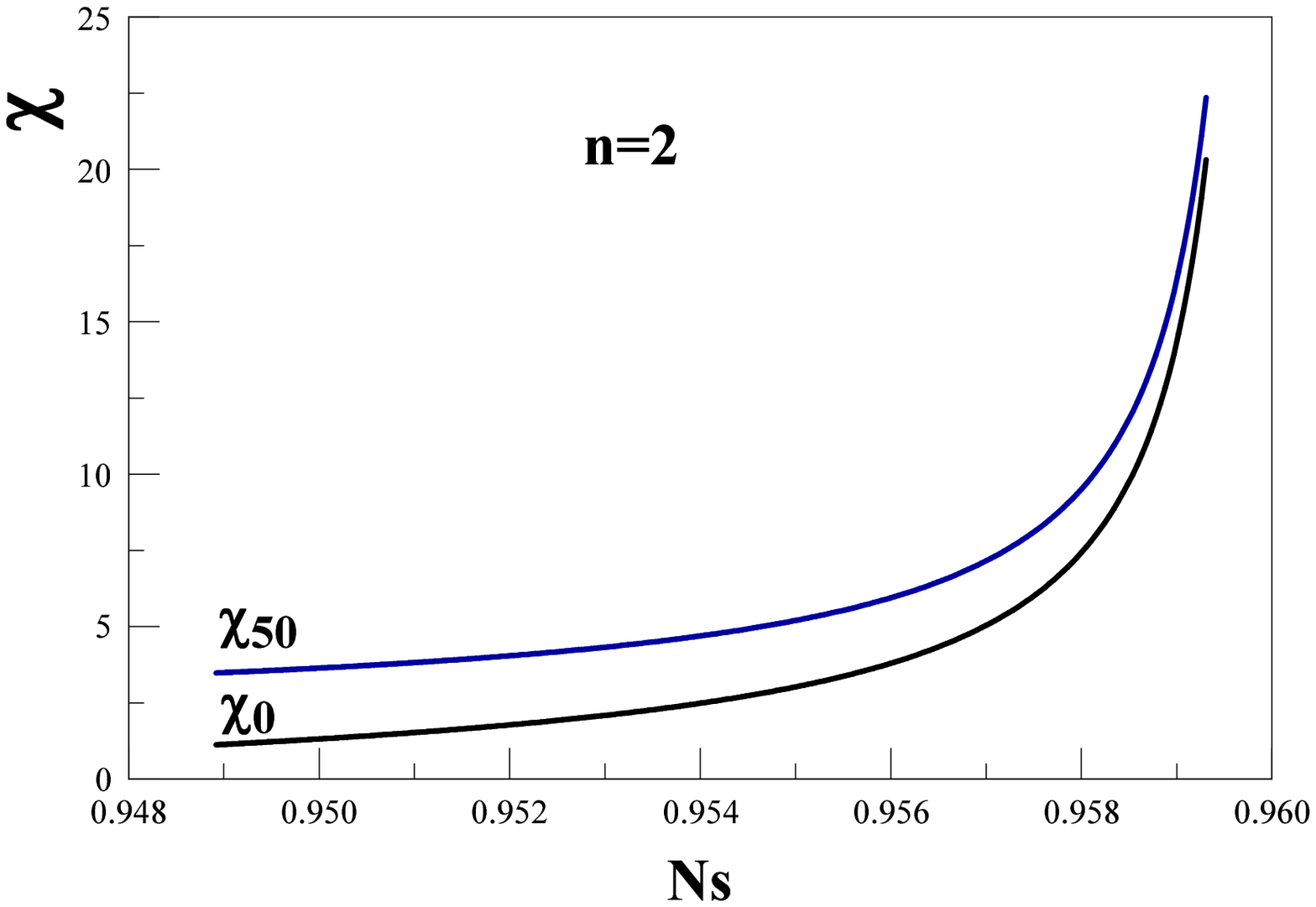}
 \includegraphics[height=2.5in,width=2.5in,keepaspectratio=true]{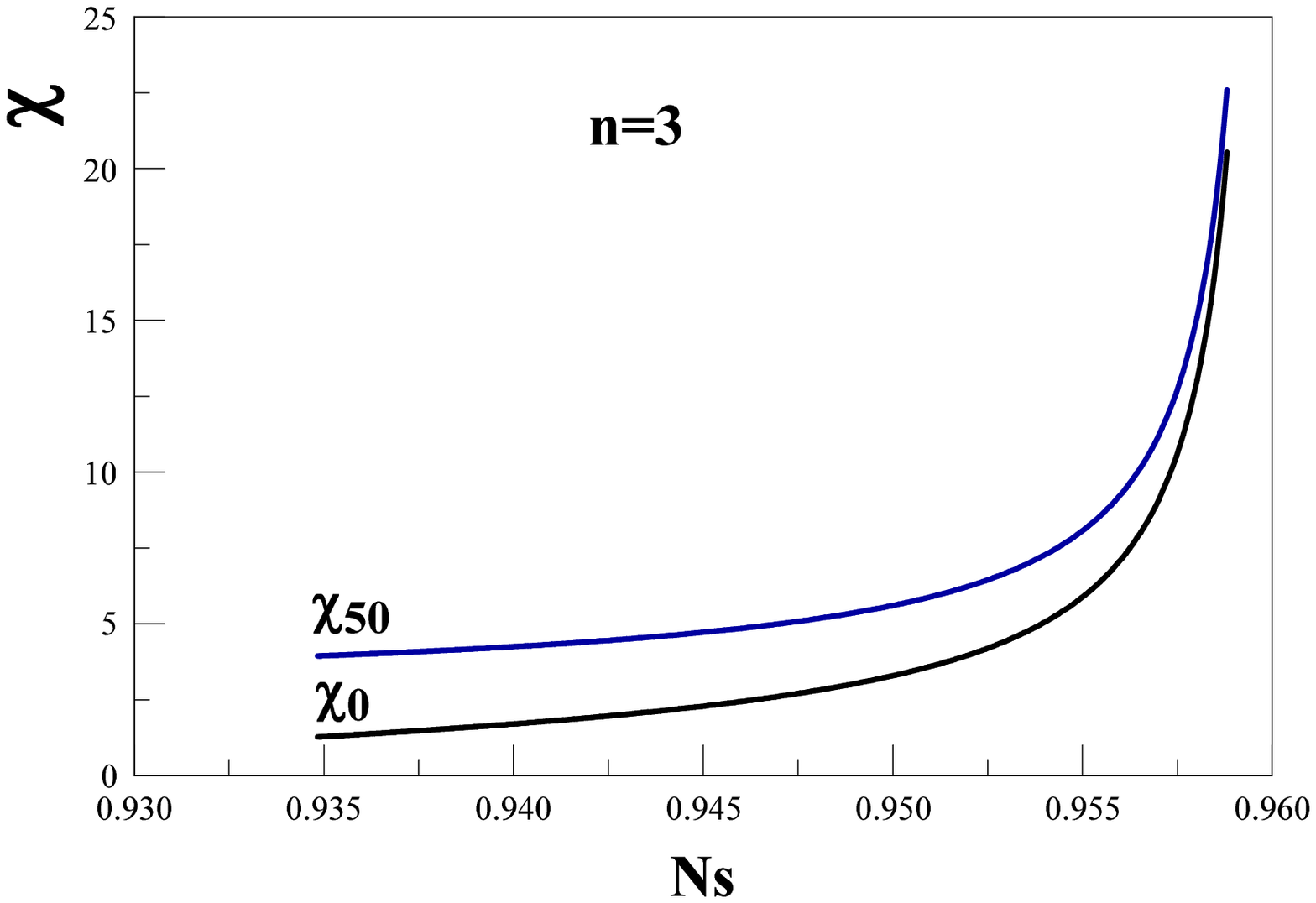}
 \includegraphics[height=2.5in,width=2.5in,keepaspectratio=true]{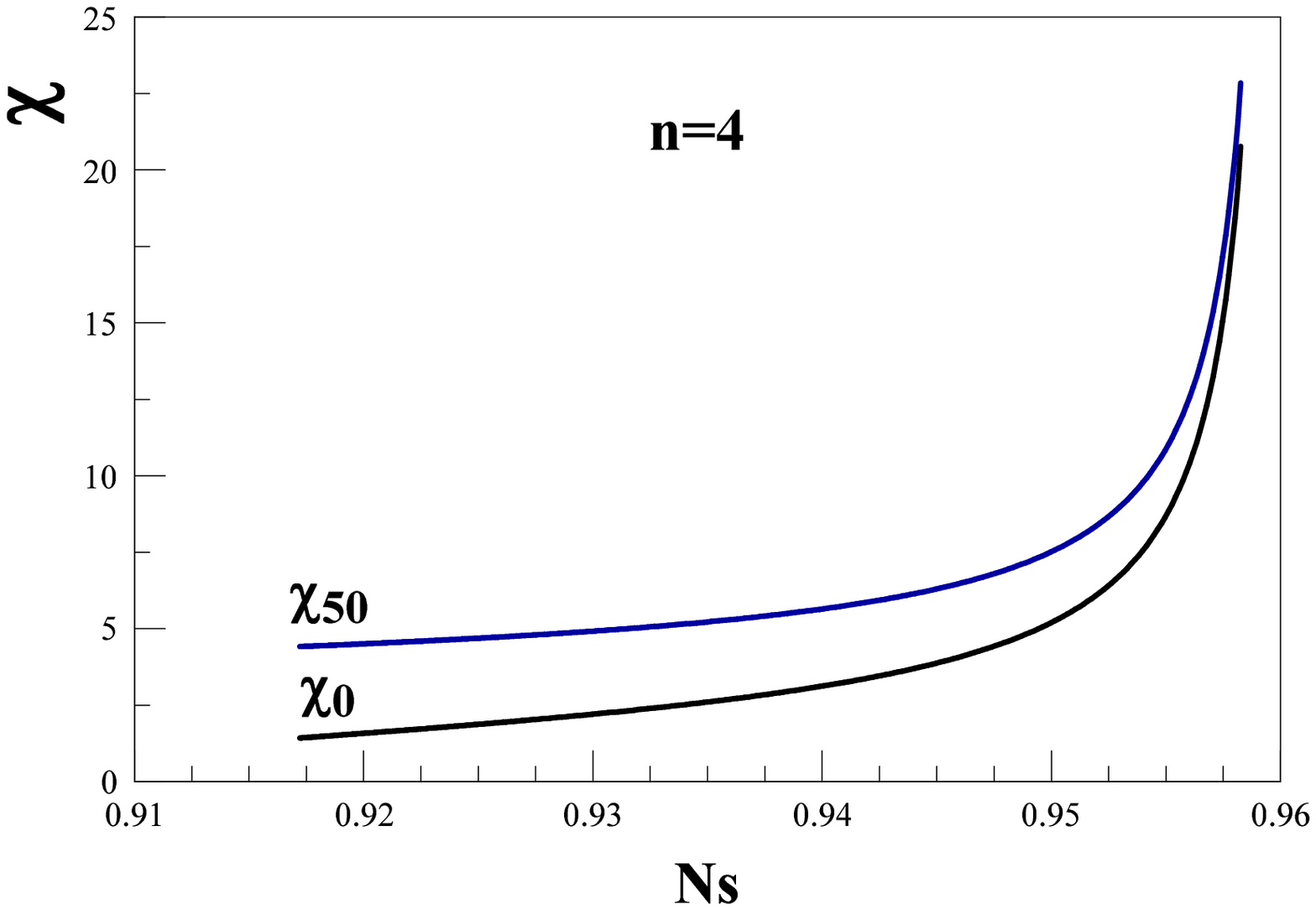}
 \caption{Reconstruction Program for broken symmetry potentials with $ N_e=50 $,
large field case $ X>1 $. $\chi_{50}\equiv \chi_c $ and $ \chi_0 $
vs. $ n_s $ for degrees of the potential
 $ n=2,3,4 $, respectively. These values of $ \chi, \; n_s $ correspond to the region $ r>0.16 $.}
 \label{fig:chinsniLF}
 \end{center}
 \end{figure}

This analysis shows that the region in field space that
corresponds to the region in $n_s$ that \emph{best fits the WMAP
data} can be systematically reconstructed in an expansion in
$ n_s-1+2/N_e $. This is yet another bonus of the $1/N_e$ expansion.
Although the above analysis can be   carried out to an arbitrary
order in $ n_s-1+2/N_e $, it is more convenient to perform a numerical
study of the region   outside from  $ X\sim 1 $ to find the values
of
 $ \chi_c $ and $ \chi_0 $ as a function of $ n_s $  for fixed values of $ n, \; N_e $. Figures
 \ref{fig:chinsniSF} and \ref{fig:chinsniLF} display  $ \chi_c, \; \chi_0 $ as a function of
 $ n_s $ with $ N_e=50 $ for different values of $ n $ for the small field region $ X<1 $ and the
large field region $ X>1 $
 respectively. The point $ X=1 $ is a degeneracy point and corresponds
 to the quadratic monomial as discussed above.

Finally, the values for the dimensionful  field $ \phi $ are given
by $ \phi_{c} = \sqrt{N_e} \; M_{Pl} \; \chi_c , \; \phi_0 =
\sqrt{N_e} \; M_{Pl} \; \chi_{0} $. Figures \ref{fig:rvnsni},
\ref{fig:chinsniSF} and \ref{fig:chinsniLF} lead to the conclusion
that for the  range of CMB parameters $ r < 0.1 $ and $ n_s \leq
0.96 $, the typical value of the {\it symmetry breaking scale} is $
\phi_0 \sim 10 \; M_{Pl} $ and the value of the inflaton field at
which cosmologically relevant wavelengths crossed the Hubble radius
during new inflation is $ \phi_{c} \sim M_{Pl} $ with a weak
dependence on $ n $. For $ 0.1 < r < 0.16 $ we have $ |\phi_{c} -
\phi_0| \sim  15 \; M_{Pl} $.

We obtain for the coupling $ g $ in the $ X \to 0 $ limit which is a strong
coupling regime [see eq.(\ref{smallX})] where $ n_s \ll 1 $,
$$
g = \left[ \frac{N_e}4 \; \left( 1 -  \frac1{n} \right) (1 - n_s) \right]^{n-1} \; .
$$
Finally, we have the $ X \to \infty $ limit which is also a strong coupling
limit [see eq.(\ref{Xgran})] where $ n_s \to 1 - (n+1)/N_e $ and we find,
\bea
&&\chi^2_0 = (4 \, n)^2 \left[\frac{N_e(1-n_s) -(n + 1)}{4 \, n (n-1) + 3 \, \left(1
+\frac1{n-2} \right)}\right]^{\frac1{n-1}} \to 0 \cr \cr
&& g = \frac{4 \, n (n-1) + 3 \, \left(1 +\frac1{n-2} \right)}{(4 \, n)^{n-1} \;
\left[N_e(1-n_s) - (n + 1)\right]} \to \infty \; .
\eea

\section{Chaotic inflation models.}

\begin{figure}[h]
\begin{center}
\includegraphics[height=3in,width=3in,keepaspectratio=true]{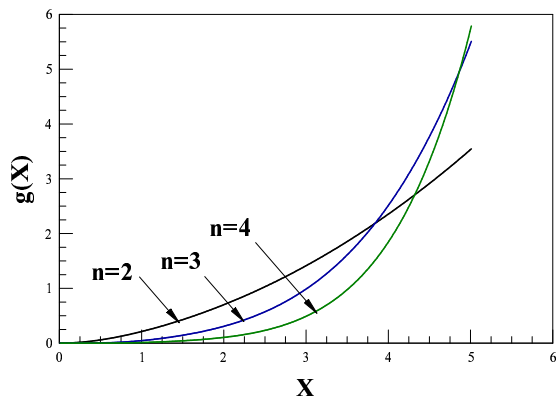}
\caption{Coupling $ g $ as a function of $ X $ for $ N_e=50 $, for
$n=2,3,4$ for chaotic inflation. $ g $ turns to be a monotonically increasing
function of $ X $. $ g $ vanishes for $ X \to 0 $ as $ \left(\frac{X}{2}\right)^{2 \, n-2} $
in sharp contrast with new inflation where $ g $ strongly increases for $ X \to 0 $.}
\label{fig:couplingscao}
\end{center}
\end{figure}

We now turn to the study of the family of chaotic inflationary
potentials given by eq. (\ref{caochi}).  Taking that the end of
inflation corresponds to $ x=0 $, the condition eq.(\ref{condi}) now
becomes 
\be 
\frac{2 \, n}{\chi^2_0} = J_n(X)  = \int^{X}_0 \frac{n
+x^{2 \, n-2}}{1+x^{2 \, n-2}}~ x~{dx}\,.\label{chis} 
\ee 
Again, this integral can be computed in closed form in terms of
hypergeometric functions \cite{gr} which can be reduced to a finite
sum of elementary functions\cite{pru}. For general values of $ X $
the integral will be studied numerically, but the small $ X $ region
can be studied by expanding the integrand in powers of $ x^{2 \,
n-2} $, with the result 
\be
1=\frac{\chi^2_0\,X^2}{4}\left[1-\frac{n-1}{n^2} \; X^{2 \, n-2}+
\mathcal{O}(X^{4 \, n-4}) \right] \; . \label{chi0cao} 
\ee 
For small $ X $ and recalling that $ X = \chi_c/\chi_0 $ this relation yields
\be \label{xi50c} |\chi_c| =2\left[1+\frac{n-1}{2 \, n^2} \; X^{2 \,
n-2}+\mathcal{O}(X^{4 \, n-4}) \right] \ee which is again, at
dominant order the relation for the quadratic monomial potential
eq.(\ref{chi50}) for $ n = 1 $. This must be the case because the
small field limit is dominated by the quadratic term in the
potential. For small fields,  $ \chi_0 \approx 2/X $ and the
coupling $ g $ vanishes as, 
\be 
g(X) \buildrel{X \to 0}\over=
\left(\frac{X}{2}\right)^{2 \, n-2} \; . 
\ee 
The coupling as a function of $ X $ is shown in fig. \ref{fig:couplingscao}.

\begin{figure}[h]
\begin{center}
\includegraphics[height=3in,width=3in,keepaspectratio=true]{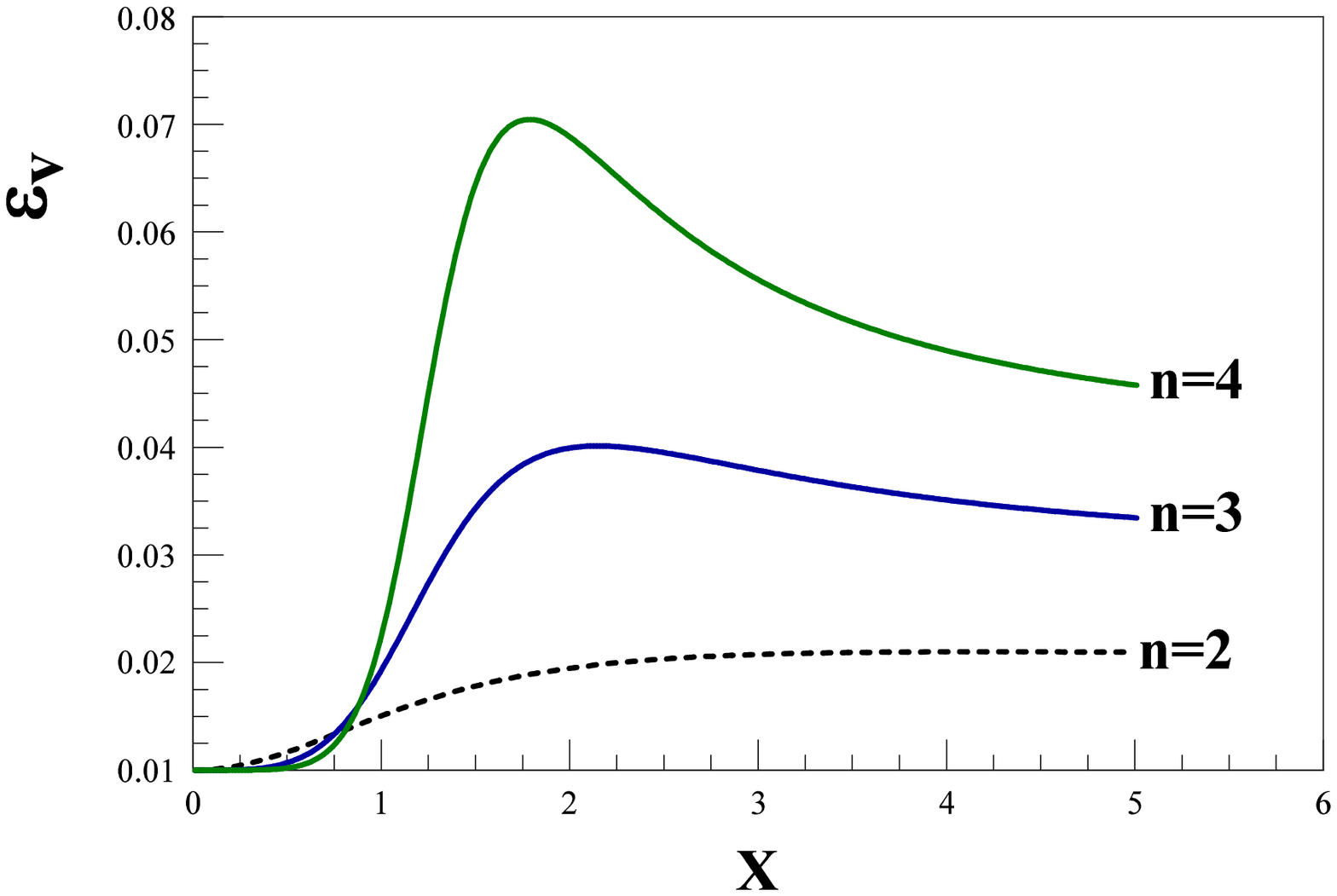}
\includegraphics[height=3in,width=3in,keepaspectratio=true]{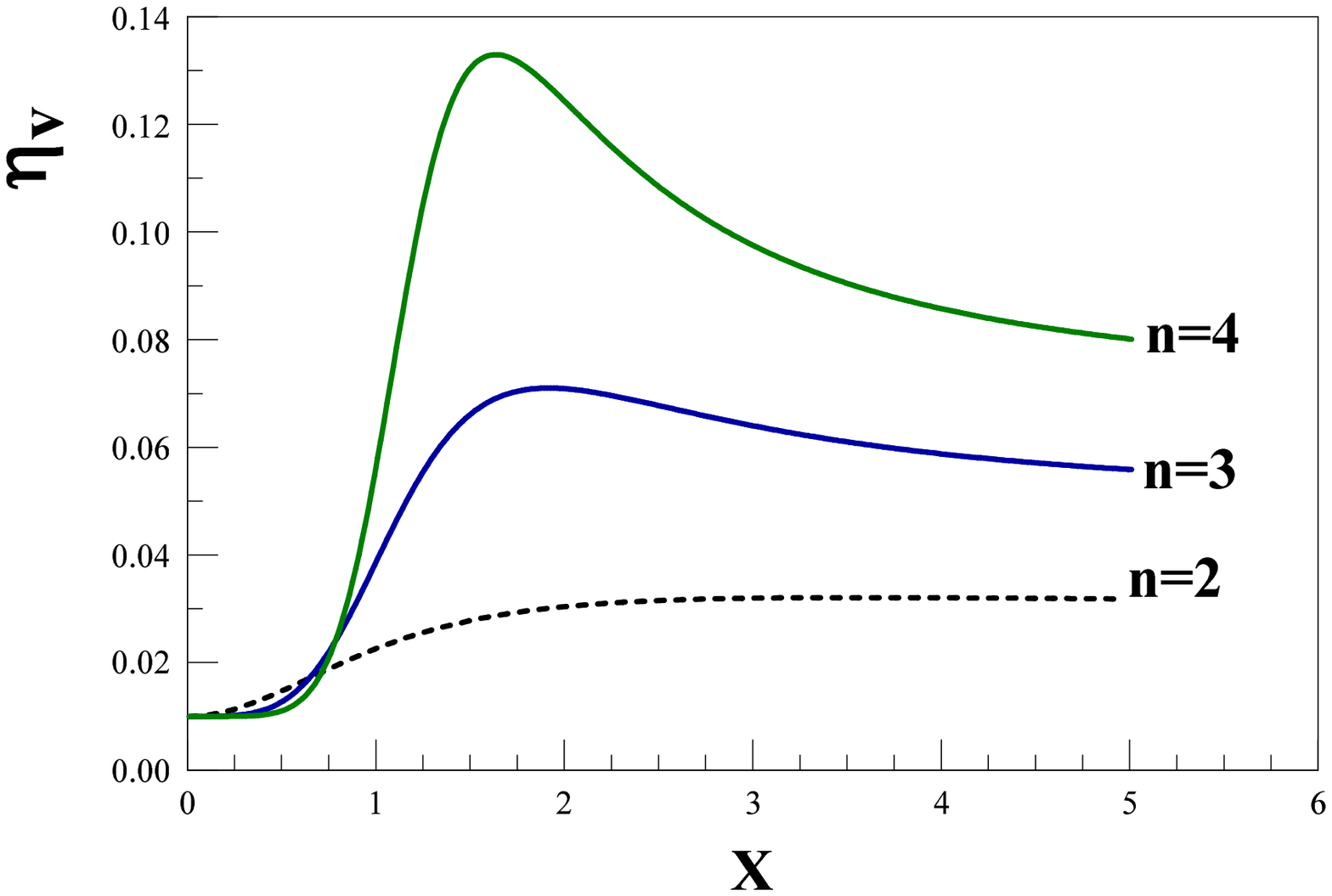}
\caption{ Left panel $ \epsilon_v $, right pannel $ \eta_v $ as a
function of $ X $ for $ N_e=50 $, for chaotic inflation with degrees of the potential  $ n=2,3,4 $.
The small $ X $ behavior is $ n $ independent.} \label{fig:epsilonvcao}
\end{center}
\end{figure}

The dependence of $ \epsilon_v, \; \eta_v $ in the full range of
$ X $ for several representative values of $ n $ is studied numerically:
these results are displayed in fig. \ref{fig:epsilonvcao}.
In the small $ X $ regime, we obtain from eqs. (\ref{etav2}), (\ref{caochi}) and (\ref{chi0cao})
the expressions,
\bea
&& \epsilon_v =
\frac{1}{2 \, N_e}\left[1+ \frac{(2 \, n-1)(n-1)}{n^2} \;
X^{2 \, n-2}+\mathcal{O}\left(X^{4 \, n-4}\right)\right] \label{evsX} \\
&&\eta_v =  \frac{1}{2 \, N_e}\left[1+ \frac{(2 \, n-1)(n^2-1)}{n^2}
X^{2 \, n-2}+\mathcal{O}\left(X^{4 \, n-4}\right)\right]
\label{etavsX} 
\eea 
As $ X\rightarrow 0 , \; \epsilon_v $ and $
\eta_v $ tend to the result from the quadratic monomial potential,
namely $ \epsilon_v=\eta_v=1/[2 \, N_e] $ as must be the case because the
quadratic term dominates the potential for $ X\ll 1 $.

\begin{figure}[h]
\begin{center}
\includegraphics[height=3.5in,width=3.5in,keepaspectratio=true]{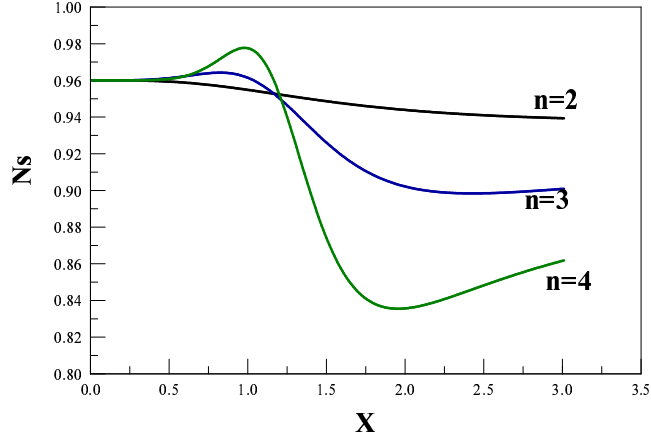}
\caption{Scalar spectral index $ n_s $ for degrees of the potential
$ n=2,3,4 $ respectively for chaotic inflation with $ N_e=50 $.  For
$ X \to 0, \; n_s $ reaches for all $ n $ the value $  n_s = 0.96 $
corresponding to the monomial potential $ m^2\phi^2/2 $. }
\label{fig:Nscao}
\end{center}
\end{figure}

\begin{figure}[h]
 \begin{center}
 \includegraphics[height=3 in,width=3 in,keepaspectratio=true]{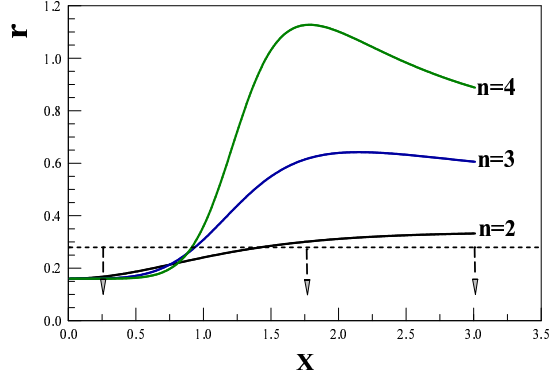}
 \caption{Tensor to scalar ratio $ r $ vs. $ X $
for degrees of the potential $ n=2,3,4 $ respectively for chaotic inflation with $ N_e=50 $.
The horizontal dashed lines with the downward arrows delimit
the region of $95\%~CL$ given by WMAP3 with no running $ r < 0.28 $ \cite{WMAP3}.}
 \label{fig:rcao}
 \end{center}
 \end{figure}

Figures \ref{fig:Nscao}, \ref{fig:rcao}  display  $ n_s, \; r $ as
functions of $ X $ for $ N_e=50 $ respectively. For $ X\rightarrow 0, \; n_s \rightarrow
0.96 $ and $ r \rightarrow 0.16 $ which are the values from the
quadratic monomial $ m^2 \; \phi^2/2 $.

For $ X\gg 1 $, the values of $ n_s, \; r $
for the monomial potentials $ \phi^{2 \, n} $ are attained asymptotically,
namely, (for $ N_e = 50 $): $ n_s-1 =   -2 \, (n+1)\times 10^{-2}  , \; r = 0.16 \, n $.

Comparing figs. \ref{fig:Nscao}, \ref{fig:rcao} to those for the
new inflation case, (figs. \ref{fig:NsNI} and \ref{fig:rni}), we note
that the range in which the chaotic family provide a
good fit to the WMAP data is \emph{very much smaller} than for new inflation.
Fig. \ref{fig:Nscao} shows that in chaotic inflation
{\it only} for $ n=2 $ the range of $ n_s $ is
allowed by the WMAP data in a fairly extensive range of values of
$ X $, whereas for $ n=3, \; 4 $ (and certainly larger), there is a
\emph{relatively small} window in field space for $ X<1 $ which satisfies the
data for $ n_s $ and $ r $ simultaneously.

The tensor to scalar ratio $ r $ in chaotic inflationary models is
{\it larger} than $ 0.16 $ for all values of $ X $, approaching asymptotically
for large $ X $ the value $ r = 0.16 \, n $  associated to the monomial potentials $ \phi^{2 \, n} $.

\begin{figure}[h]
 \begin{center}
 \includegraphics[height=3.5 in,width=3.5 in,keepaspectratio=true]{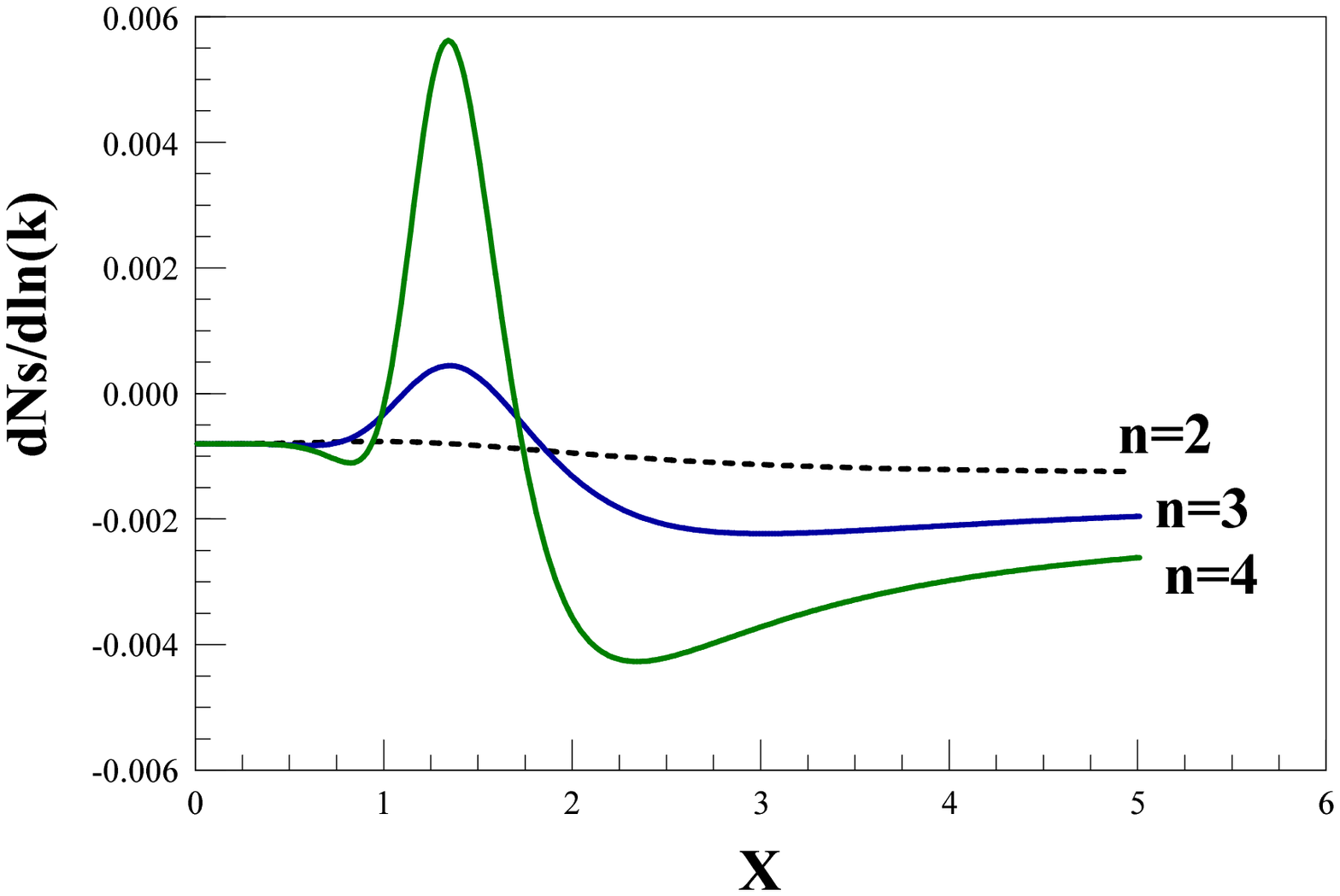}
 \caption{Running of the scalar index $ dn_s/d\ln k $ vs. $ X $ for degrees of the potential
   $ n=2, \; 3, \; 4 $  respectively for chaotic inflation with $ N_e=50 $.
The  $ X \to 0 $ behavior is $ n $ independent. $ dn_s/d\ln k $ features a maximun value
that gets stronger with increasing $ n $. For chaotic inflation $ dn_s/d\ln k $ takes negative
as well as positive values, in contrast with new inflation where  $ dn_s/d\ln k < 0$.}
 \label{fig:dnscao}
 \end{center}
 \end{figure}

 Fig. \ref{fig:dnscao} displays $ dn_s/d\ln k $ as a function of
 $ X $, while the running is again negligible, it is {\it strikingly
 different} from the new inflation case. \emph{Again} this figure, in
 combination with those for $ n_s $ and $ r $ as functions of $ X $
 distinctly shows that {\bf only} $ n=2 $ in chaotic inflation is compatible with the
 bounds from the WMAP data, while for $ n=3,4 $ only a {\it small window}
 for $ X<1 $ is allowed by the data. We see from fig. \ref{fig:dnscao} that
$ dn_s/d\ln k $ takes negative as well as positive values for chaotic inflation,
in contrast with new inflation where  $ dn_s/d\ln k < 0 $.

\begin{figure}[h]
 \begin{center}
 \includegraphics[height=3 in,width=3 in,keepaspectratio=true]{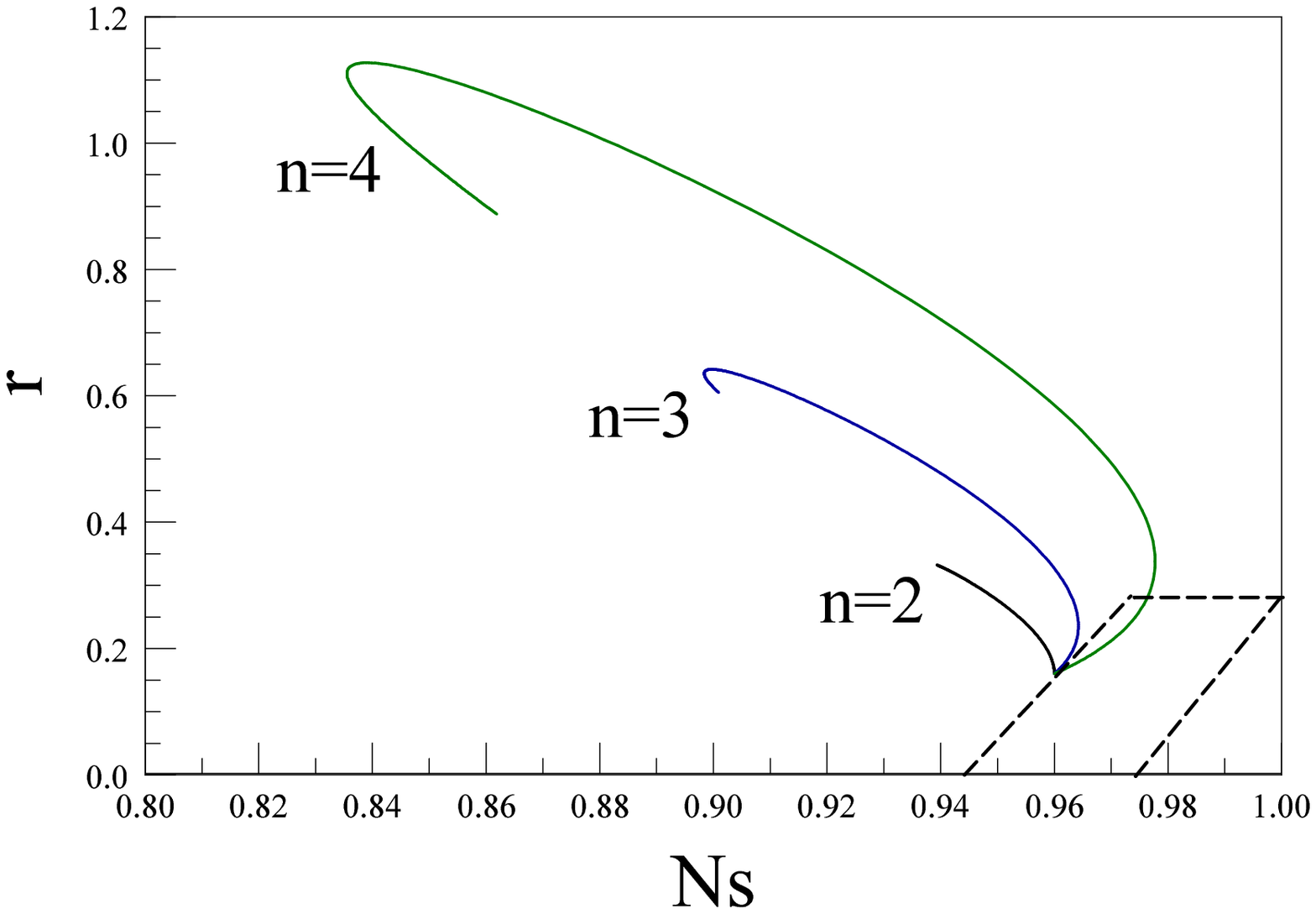}
 \caption{Tensor to scalar ratio $ r $ vs. $ n_s $
  for degrees of the potential  $ n=2, \; 3, \; 4 $ respectively
for chaotic inflation with $ N_e=50 $.
The range of $ 95\%~CL $ as determined by WMAP3 \cite{WMAP3} is within
  the tilted box delimited by :
$ r < 0.28, \;  0.942+0.12 \, r \leq n_s \leq 0.974+0.12 \, r $, see
eq.(\ref{ley}).}
 \label{fig:rvsncao}
 \end{center}
 \end{figure}

The fact that the combined bounds on $ n_s, \; r $ and $ dn_s/d\ln k
$ from the WMAP3 data \cite{WMAP3} provide much more stringent
constraints on chaotic models is best captured by displaying $ r $
as a function of $ n_s $ in fig. \ref{fig:rvsncao}. The region
allowed by the WMAP data lies within the tilted box delimited by the
vertical and horizontal dashed lines that represent the $ 95\% CL $
band \cite{WMAP3}.

\begin{figure}[h]
 \begin{center}
 \includegraphics[height=3.5 in,width=3.5 in,keepaspectratio=true]{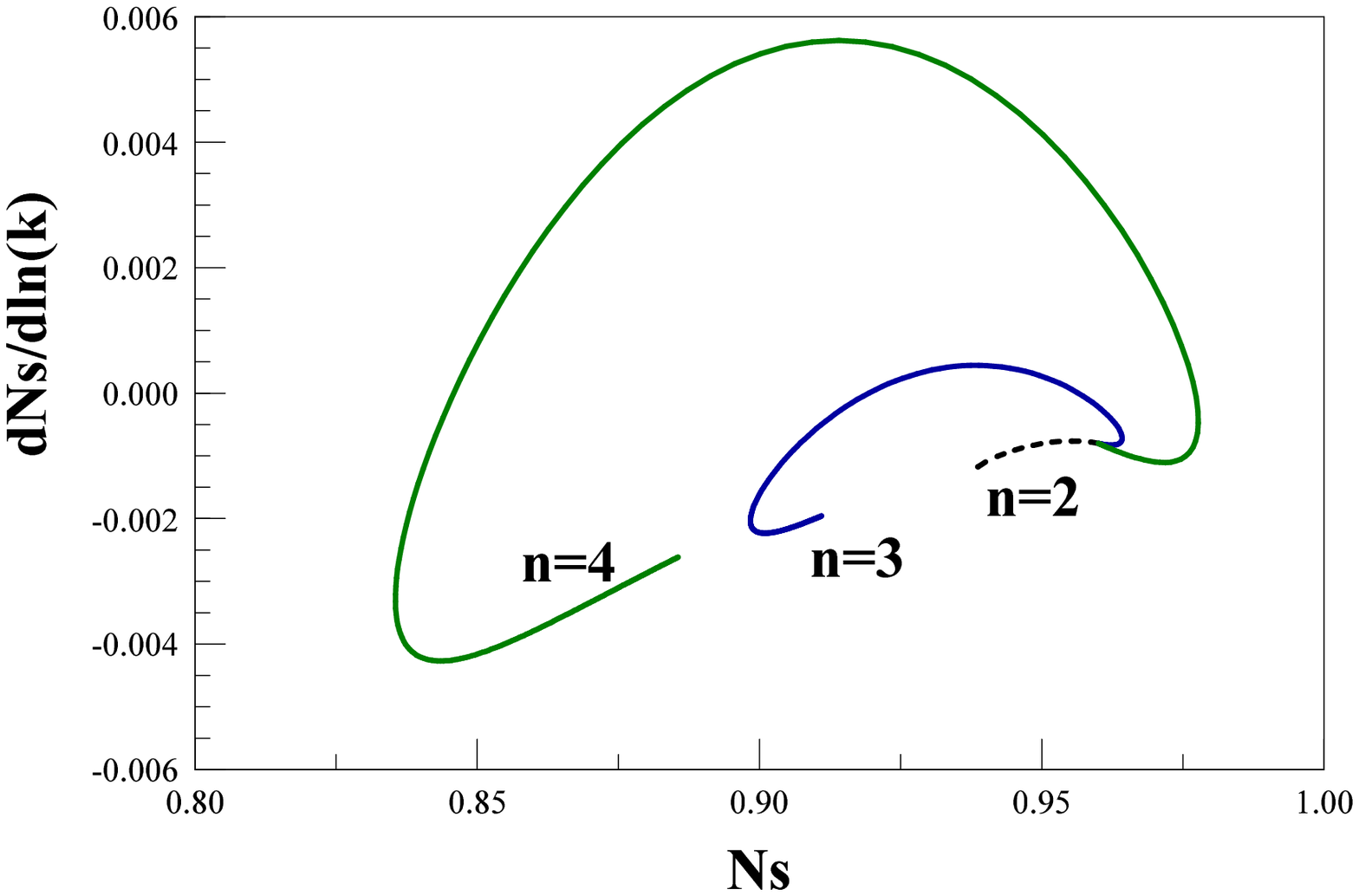}
 \caption{Running of the scalar index $ dn_s/d\ln k $ vs. $ n_s $  for degrees of the potential
   $ n=2, \, 3, \, 4 $ respectively for chaotic inflation with $ N_e=50 $. }
 \label{fig:ndnscao}
 \end{center}
 \end{figure}

 A complementary assessment of the
 allowed region for this family of effective field theories is
 shown in fig. \ref{fig:ndnscao} which distinctly shows that {\bf only}
 the $ n=2 $ case of chaotic inflation is allowed by the WMAP3 data.

\subsection{Field reconstruction  }

The reconstruction program proceeds in the same manner as in the
case of new inflation: the first step is to obtain $ \chi_0(X) $
from eq.(\ref{chis}). Then $ \epsilon_v $ and $ \eta_v $ are
obtained as a function of $ X $ which yields $ n_s(X) $. Inverting
this relation we find $ X=X(n_s) $ and finally $ \chi_c(n_s) =
\chi_0 \; X(n_s) $. While this program must be carried out
numerically, we can gain important insight by focusing on the small
$ X $ region and using eq.(\ref{chi0cao}).

From eqs. (\ref{ns}), (\ref{evsX}) and (\ref{etavsX}) we find 
\be
n_s-1+\frac{2}{N_e} = X^{2 \, n-2} \; \frac{(2 \,
n-1)(n-1)(n-2)}{n^2\,N_e}+\mathcal{O}\Big(X^{4n-4}\Big)\label{expacao}
\ee 
As $ X \rightarrow 0 $ it follows that $ n_s \rightarrow 1-2/N_e
$ which is the value for the scalar index for the quadratic monomial
potential $ m^2\phi^2/2 $. However, for $n>2$ this limit is
approached from above, namely for $ n>2 $ it follows that
$ n_s>1-2/N_e $. The small $ X $ region corresponds to small departures
of $ n_s $ from the value determined by the quadratic monomial $
1-2/N_e $ but always larger than this value for $ n>2 $. In the small
field limit we reconstruct the value of $ \chi_c $ in an expansion in
$ n_s-1+2/N_e $. The leading order in this expansion is obtained by
combining eqs. (\ref{xi50c}) and (\ref{expacao}), we obtain 
\be
\label{chifif} |\chi_c|
=2\Bigg[1+\frac{\big(n_s-1+\frac{2}{N_e}\big)N_e}{2 \, (2 \, n-1) \,
(n-2)}\Bigg]+ {\cal O}\left( \left[n_s-1+2/N_e \right]^2\right) 
\ee
Obviously, this leading order term is singular at $ n=2 $, this is a
consequence of the result eq.(\ref{expacao}) which entails that for
$ n=2 $ the expansion must be pursued to higher order, up to $
X^{4n-4} $.

We find from eqs. (\ref{ns}), (\ref{evsX}) and (\ref{etavsX}) for $
n = 2 $, 
\be \label{neq2} 
n_s-1+\frac{2}{N_e} = - \frac{17}{24 \,
N_e} \; X^4 + \mathcal{O}\Big(X^6\Big) \quad , \quad n=2 \; , 
\ee
  therefore, 
\be\label{chin2} 
|\chi_c| =2\Bigg[1+\sqrt{\frac{3
\, N_e}{136}\left(1 -\frac2{N_e}-n_s \right)}\,\Bigg] + {\cal
O}\left(n_s-1+\frac2{N_e} \right)\quad , \quad n=2 \; . 
\ee 
We see that the derivative of $ \chi_c $ with respect to $ n_s $ is
singular for $ n=2 $ at $ n_s=1-\frac2{N_e} $. We note that for
$ n=2 $ there is a sign change with respect to the cases $ n>2 $ and
$ n_s-1+2/N_e \leq 0 $ as determined by eq. (\ref{neq2}).

\begin{figure}[h]
 \begin{center}
 \includegraphics[height=3.5 in,width=3.5 in,keepaspectratio=true]{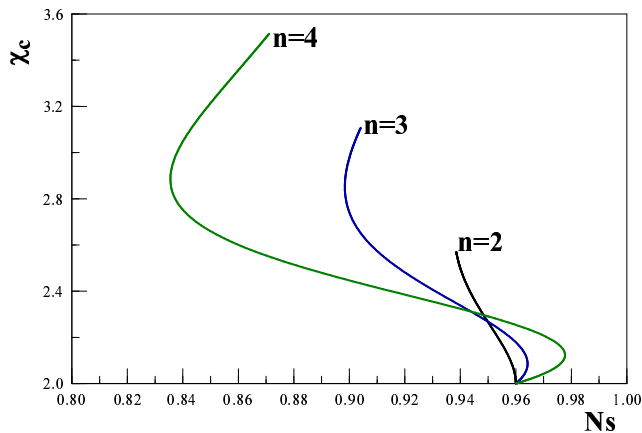}
 \caption{Reconstruction program  for chaotic inflation with $ N_e=50 $,
$ \chi_c $ vs. $ n_s $ for $ n=2, \, 3, \, 4 $ respectively. }
 \label{fig:chinscao}
 \end{center}
 \end{figure}

Fig. \ref{fig:chinscao} shows $ \chi_c $ as a function of $ n_s $
for $ n=2, \; 3, \;4 $ for $ N_e=50 $. The case $ n=2 $ clearly
shows the singularity in the derivative $ \partial  \chi_c/
\partial n_s $ at $ n_s = 1-2/N_e = 0.96 $ [see eq.(\ref{chin2})].

Combining fig. \ref{fig:chinscao} with fig. \ref{fig:rvsncao} it is
clear that there is a small window in field space within which
chaotic models provide a good fit to the WMAP3 data, for $N_e=50$ we
find:
\bea \label{regiochi50} 
&& n =2 ~:   ~~0.95 \lesssim n_s \leq 0.960
~,~2.0 \leq |\chi_c| \lesssim 2.25  \nonumber \\&& n =3~: ~~0.96
\leq n_s  \lesssim 0.965 ~,~2.0 \leq |\chi_c| \lesssim 2.15 \nonumber \\
&& n =4~:   ~~0.96 \leq n_s \leq 0.975 ~,~2.0 \leq |\chi_c| \lesssim
2.10  \; .
\eea 
Restoring the dimensions via eq. (\ref{chi}) these
values translate into a narrow region of width $\Delta \phi \lesssim
1.5\,M_{Pl}$ around the scale $|\phi_c| \sim 15~M_{Pl}$.

Therefore, the joint analysis for $ n_s, \; r, \; dn_s/d\ln k $
distinctly reveals that: (i) chaotic models favor {\it larger}
values of $ r $ thus, larger tensor amplitudes, and (ii) chaotic models feature
{\it smaller} regions in field space consistent with the CMB and
large scale structure data.  {\it Only} the case $ n=2 $ features a
larger region of consistency with the combined WMAP3 data.

\section{Conclusions}

The fact that current  observations of the CMB and LSS are already
placing constraints  on inflationary models which will undoubtedly
become more stringent with forthcoming observations, motivates a
study of the predictions for the CMB power spectra from different
inflationary scenarios. We perform a systematic study of
\emph{families} of single field new and chaotic inflation slow roll
models characterized by effective field theories with potentials of
the form
\bea
V(\phi) &=&  V_0 -\frac{1}{2} \; m^2 \; \phi^2 +
\frac{\lambda}{2 \, n}\; \phi^{2n} \quad , \quad {\rm broken ~ symmetry}
\label{nuev}\\V(\phi) & = & \frac{1}{2} \;  m^2 \; \phi^2 +
\frac{\lambda}{2 \, n} \; \phi^{2 \, n} \quad , \quad {\rm unbroken ~
symmetry}\label{chau} \; .
\eea
Unlike the approach followed in \cite{peiris,WMAP3} based on the
inflationary flow equations\cite{hoki}, or more recent
studies which focused on specific inflationary
models \cite{salama}, or on statistical analysis of models \cite{kinkol},
we implement an expansion in $ 1/N_e $ where $ N_e \sim
50 $ is the number of e-folds before the end of inflation when
wavelengths of cosmological relevance today cross the Hubble radius
during inflation. We provide an analysis of the dependence of CMB
observables ($ n_s, \; r $ and $ dn_s/d\ln k $) with $ n $ and
establish the region in field space within which these families
provide a good agreement with the WMAP3 data combined with large
scale surveys.

For new inflation models with potentials eq.(\ref{nuev}) there are two distinct
regions corresponding to values of the inflaton field  smaller
(small field) or larger (large field) than the symmetry breaking
scale. For this family  we find a {\bf wide range} in the $(n_s,r)$ 
plane in which the different members $ n=2, \; 3, \; 4...$ are
allowed by the data both for small and large fields with negligible
running of the scalar index:
$$
-4 \; (n+1)\times 10^{-4} \leq  dn_s/d\ln k   \leq -2\times 10^{-4}  \; .
$$
For $ N_e=50 $ the values $ n_s=0.96,r=0.16 $ which are those
determined by the simple monomial potential $ m^2 \; \phi^2/2 $
determine a divide and a degeneracy point in the field and parameter
space. Small field regions yield $ r< 0.16 $ while large field
regions correspond to $ r>0.16 $.

The $ 1/N_e $ expansion also provides a powerful tool to
implement a {\it reconstruction program} that allows to extract the
value of the field $ N_e $ e-folds before the end of inflation, and in
the case of new inflationary models, the symmetry breaking scale.

We find that the region of field space favored by the WMAP3 data can
be explored in a systematic expansion in $n_s-1+2/N_e$. An analytic
and numerical study of this region lead us to conclude that if
forthcoming data on tensor modes favors $ r<0.16 $ then
new inflation {\bf is favored}, and we {\bf predict} for $ r < 0.1 $ that
(i) the  {\bf symmetry breaking scale} is
$$
\phi_0 \sim 10~M_{Pl} \; ,
$$
and (ii) the value of the field when cosmologically relevant
wavelengths cross the Hubble radius is $ |\phi_c|\sim M_{Pl} $ .

The family of chaotic inflationary models characterized by the
potentials eq.(\ref{chau}) feature tensor to scalar ratios $ r \geq
0.16 $ (for $ N_e=50 $), with the minimum, $ r=0.16 $ obtained in
the limit of small inflaton amplitude and corresponds to the
monomial potential $ m^2 \; \phi^2/2 $ which is again a degeneracy
point for this family of models.

The combined marginalized data from WMAP3 \cite{WMAP3} yields a very
small window within which chaotic models are allowed by the data,
the largest region of overlap with the $ (r,n_s) $ WMAP3 data
corresponds to $ n=2 $ and the width of the region decreases with
larger $n$. The typical scale of the field at Hubble crossing for
these models is $|\phi_c|\sim 15~M_{Pl}$ (for $N_e = 50$). Some
small regions in field space consistent with the WMAP3 data feature
peaks in the running of the scalar index but in the region
consistent with the WMAP3 data in chaotic inflation the running is
again negligible. If future observations determine a tensor to
scalar ratio $ r<0.16 $, such bound will, all  by itself, {\bf rule
out} the  large family of chaotic inflationary models of the form
(\ref{chau}) for any $ n $.

\begin{acknowledgments}    D. B. and C. M. Ho thank the US NSF for support under grant
PHY-0242134,  LPTHE and the Observatoire de Paris and LERMA for
hospitality during this work. D.B. thanks A. Kosowsky for
illuminating conversations.
\end{acknowledgments}


\begin{thebibliography}{99}
\bibitem{infla} D. Kazanas, Astrophys. J \textbf{241}, L59 (1980).
A. Guth, Phys. Rev. \textbf{D23}, 347 (1981); astro-ph/0404546 (2004).
K. Sato, Mon. Not. R. Astron. Soc. \textbf{195}, 467 (1981).

\bibitem{libros} E. W. Kolb and M. S. Turner, {\em The Early Universe}
Addison Wesley, Redwood City, C.A. 1990.
P. Coles and F. Lucchin, {\em Cosmology}, John Wiley, Chichester,
1995.
A. R. Liddle and D. H. Lyth,
{\em Cosmological Inflation and Large Scale Structure}, Cambridge
University Press, 2000.
S. Dodelson, {\em Modern Cosmology}, Academic Press, 2003.
D. H. Lyth , A. Riotto, Phys. Rept. \textbf{314}, 1 (1999).

\bibitem{fluc} V. F. Mukhanov , G. V. Chibisov, Soviet Phys.
JETP Lett. \textbf{33}, 532 (1981); V. F. Mukhanov, H. A. Feldman ,
R. H. Brandenberger, Phys. Rept. \textbf{215}, 203 (1992).
S. W. Hawking, Phys. Lett. \textbf{B115}, 295 (1982).
A. H. Guth , S. Y. Pi, Phys. Rev. Lett. \textbf{49}, 1110 (1982).
A. A. Starobinsky, Phys. Lett. \textbf{B117}, 175 (1982).
J. M. Bardeen, P. J. Steinhardt , M. S. Turner, Phys. Rev.
\textbf{D28}, 679 (1983).

\bibitem{WMAPa} C. L. Bennett \emph{et.al.} (WMAP collaboration),
Ap. J. Suppl. \textbf{148}, 1 (2003).
A. Kogut  \emph{et.al.} (WMAP collaboration), Ap. J. Suppl.
\textbf{148}, 161 (2003).
D. N. Spergel \emph{et. al. }(WMAP collaboration), Ap. J. Suppl.
\textbf{148}, 175 (2003).

\bibitem{peiris} H. V. Peiris \emph{et.al.} (WMAP collaboration), Ap. J.
Suppl.\textbf{148}, 213 (2003).

\bibitem{WMAP3} D. N. Spergel \emph{et. al. }(WMAP collaboration), astro-ph/0603449.

\bibitem{WMAP3b} L. Page, \emph{et. al. }(WMAP collaboration), astro-ph/0603450.
G. Hinshaw,\emph{et. al. }(WMAP collaboration), astro-ph/0603451.
N. Jarosik, \emph{et. al. }(WMAP collaboration),
astro-ph/0603452.


\bibitem{SDSS} M. Tegmark \emph{et.al.} Phys. Rev
\textbf{D69},103501, (2004)
D. J. Eisenstein \emph{et.al.} ApJ. \textbf{633}, 560 (2005)
U. Seljak et al.,  Phys. Rev. D71, 103515 (2005).

\bibitem{2dF} A. G. S\'anchez \emph{et. al.}, Mon. Not. Roy. Astron. Soc. {\bf 366},
189 (2006). S. Cole \emph{et.al.} Mon. Not. Roy. Astron. Soc. {\bf 362}, 505
(2005).

\bibitem{sanchez} A. G. Sanchez and C. M. Baugh, astro-ph/0612743.

\bibitem{hoki} M. B. Hoffman and M. S. Turner, Phys. Rev.
\textbf{64}, 023506 (2001). W. H. Kinney, Phys. Rev. \textbf{D66},083508 (2002).

\bibitem{salama} C. Savage, K. Freese and W. H. Kinney, Phys. Rev. \textbf{D 74}, 123511
(2006); L. Alabidi and D. H. Lyth, astro-ph/0510441,
astro-ph/0603539; J. Martin and C. Ringeval, JCAP \textbf{0608}
(2006) 009.

\bibitem{kinkol} W. H. Kinney, E. W. Kolb, A. Melchiorri and A.
Riotto, Phys.Rev. D74 (2006) 023502.

\bibitem{fipele} F. Finelli, M. Rianna, N. Mandolesi, JCAP
\textbf{0612},  006 (2006). H. Peiris, R. Easther, JCAP \textbf{0610},  017
(2006); JCAP \textbf{0607}, 002 (2006); R. Easther, H. Peiris, JCAP
\textbf{0609}, 010 (2006). S. M. Leach, A. R Liddle, Phys.Rev. \textbf{D68}, 123508
(2003); Mon.Not.Roy.Astron.Soc. \textbf{341}, 1151 (2003).

 \bibitem{1sN}
D. Boyanovsky, H. J. de Vega and N. G. Sanchez, Phys. Rev.
\textbf{D73}, 023008 (2006). See also ref. \cite{mangano}.

\bibitem{reco} J. Lidsey, A.
Liddle, E. Kolb, E. Copeland, T. Barreiro and M. Abney, Rev. of Mod.
Phys. {\bf 69}, 373, (1997).

\bibitem{Ne} W. H. Kinney, A. Riotto, JCAP \textbf{0603}, 0.11
(2006); M. Giovannini, arXiv:astro-ph/0703730.

\bibitem{ciri} D. Cirigliano, H. J. de Vega and N. G. Sanchez, Phys.
Rev. \textbf{D71}, 103518 (2005);

\bibitem{pre} H. J. de Vega
and N. G. Sanchez, Phys. Rev. \textbf{D74}, 063519 (2006).

\bibitem{heclast} C. Destri, H. J. de Vega, N. G. Sanchez, arXiv:astro-ph/0703417

\bibitem{barrow} A. R. Liddle, P. Parsons , J. D. Barrow, Phys.
Rev. \textbf{D50}, 7222 (1994).

\bibitem{mangano} G. Mangano, G. Miele, C. Stornaiolo,  Mod.Phys.Lett.A10, 1977
(1995).

\bibitem{gr}  I. S. Gradshteyn and I. M. Ryshik, Table of Integrals, Series
and Products, Academic Press, 1980.

\bibitem{pru} A. P. Prudnikov, Yu. A. Brichkov, O, I. Marichev, Integrals and Series,
vol. 3, Nauka, Moscow, 1986.

\end{thebibliography}
\end{document}